\documentclass[cameraready]{Interspeech}

\title{EvoTSE: Evolving Enrollment for Target Speaker Extraction}

\author[affiliation={1}]{Zikai}{Liu}
\author[affiliation={1}]{Ziqian}{Wang}
\author[affiliation={1}]{Xingchen}{Li}
\author[affiliation={1}]{Yike}{Zhu}
\author[affiliation={2}]{Shuai}{Wang}
\author[affiliation={3}]{Longshuai}{Xiao}
\author[affiliation={1},correspondingauthor]{Lei}{Xie}

\address{
    $^1$ Audio, Speech and Language Processing Group (ASLP@NPU),\\ School of Software, Northwestern Polytechnical University, Xi'an, China \\
    $^2$ Nanjing University, China \\
    $^3$ Huawei Technologies Co., Ltd., China
}

\email{liuzikai@mail.nwpu.edu.cn, lxie@nwpu.edu.cn}


\newcommand{\drawwaveform}[4]{
    \foreach \i in {1,...,30} {
        \pgfmathsetmacro{\h}{0.28 * (0.5 + 0.5*sin(\i*20 + #1*50)) * (0.7 + 0.3*rnd)}
        \draw[#2, line width=0.7pt, line cap=round] 
            ($(#3) + (\i*0.06 - 0.9, \h)$) -- ($(#3) + (\i*0.06 - 0.9, -\h)$);
    }
}
\newcommand{\drawwaveformshort}[4]{
    \foreach \i in {1,...,18} {
        \pgfmathsetmacro{\h}{0.22 * (0.5 + 0.5*sin(\i*25 + #1*70)) * #4 * (0.7 + 0.3*rnd)}
        \draw[#2, line width=0.6pt, line cap=round] 
            ([xshift=\i*0.05cm-0.45cm]#3) -- +(0,\h) -- +(0,-\h);
    }
}

\usepackage{comment}
\usepackage{subcaption}
\usepackage{pgfplots}
\usepackage{booktabs}
\usepackage{tabularx}
\usepackage{makecell} 
\usepackage{multirow}
\usepackage{makecell, multirow, booktabs, amssymb}
\usetikzlibrary{arrows.meta, positioning, calc, shapes.geometric}
\pgfplotsset{compat=1.18}

\begin{document}

\maketitle

\begin{abstract}

    Target Speaker Extraction (TSE) aims to isolate a specific speaker’s voice from a mixture, guided by a pre-recorded enrollment. While TSE bypasses the global permutation ambiguity of blind source separation, it remains vulnerable to speaker confusion, where models mistakenly extract the interfering speaker. Furthermore, conventional TSE relies on static inference pipeline, where performance is limited by the quality of the fixed enrollment. To overcome these limitations, we propose EvoTSE, an evolving TSE framework in which the enrollment is continuously updated through reliability-filtered retrieval over high-confidence historical estimates. This mechanism reduces speaker confusion and relaxes the quality requirements for pre-recorded enrollment without relying on additional annotated data. Experiments across multiple benchmarks demonstrate that EvoTSE achieves consistent improvements, especially when evaluated on out-of-domain (OOD) scenarios. Our code and checkpoints are available\footnote{\url{https://github.com/IiuZiKai/Evo_TSE}}.

\end{abstract}

\section{Introduction}
Target speaker extraction aims to isolate a desired voice from multi-talker mixtures using a reference enrollment. Despite recent progress, practical deployment is fundamentally limited by two challenges. First, speaker confusion remains a critical failure mode, where models incorrectly track interfering speakers that exhibit similar vocal characteristics or emotional intensities to the target~\cite{TargetConfusion, xsep_chunkloss}. Second, conventional frameworks rely on a static inference pipeline with a fixed enrollment signal. This creates a static-dynamic mismatch during long-duration processing: the target speaker's voice undergoes intrinsic acoustic drift due to emotional changes or varying vocal efforts. Consequently, a static enrollment fails to represent the time-varying acoustic characteristics, leading to significant performance degradation in OOD scenarios~\cite{ortse}.

Existing research has focused predominantly on architectural refinements. From early frameworks~\cite{voicefilter, Wang2018DeepEN, speakerbeam, HeShulin20} to recent models such as USEF-TSE~\cite{USEFTSE} and X-TF-GridNet~\cite{HAO2024102550}, substantial progress has been made in enhancing identity discriminability, particularly in challenging same-gender mixtures. However, these backbones still rely on a fixed enrollment, which is inherently fragile when the target speaker’s acoustic characteristics shift or when an interfering speaker aligns closely with the initial fixed enrollment. Consequently, conventional independent mapping remains inadequate to handle the cumulative identity drift and severe speaker confusion found in continuous long-duration scenarios.

Another research direction for mitigating speaker confusion focuses on refining the target speaker's enrollment. Theoretically, the accuracy of the extraction process is highly dependent on the quality and representativeness of the enrollment. If an enrollment is acoustically or emotionally closer to an interferer than to the target, the system becomes prone to identity mismatches~\cite{ortse, compareenroll}. Existing studies in this domain have primarily targeted two objectives: improving the robustness of the speaker encoder against noisy or overlapping enrollment signals~\cite{EnrollAug, Veluri2024LookOT, GhaneEnroll} and deriving enrollment from the inference session itself. Regarding the latter, some frameworks derive reference information directly from non-overlapping segments of the mixture~\cite{SADenroll, Nonoverlapp} or utilize iterative refinement strategies to adapt the speaker representation across multiple extraction stages~\cite{DPRNNIRA, SpEx++}. However, while these methodologies significantly enhance the model's tolerance to enrollment noise or enrollment scarcity~\cite{shortenroll}, they rarely address the fundamental problem of selecting an optimal, context-aware enrollment to resolve speaker confusion. Most current approaches aim for a stable average representation of the speaker’s identity, which remains insufficient for tracking the dynamic acoustic variations that lead to speaker confusion in non-stationary scenarios.

To address these limitations, we propose a framework that transitions from static processing to an evolving TSE pipeline. This approach treats inference over long-duration audio as a continuous process, enabling the system to explicitly utilize historical context to track the target speaker's intrinsic acoustic drift as vocal characteristics evolve. Inspired by Retrieval-Augmented Generation (RAG)~\cite{RAG}, we adaptively refine the reference information by dynamically retrieving the most acoustically relevant enrollments from high-confidence historical estimates. EvoTSE reduces speaker confusion when interferers resemble the enrollment signal and relaxes the quality requirements for pre-recorded enrollment audio. 

Our contributions are as follows:
\begin{itemize} 
\item We reformulate the static TSE task into an evolving pipeline that explicitly leverages historical context.
\item We propose EvoTSE with reliability filtering to adaptively update speaker cues, along with an artifact-aware two-stage training strategy that improves robustness to the artifact-containing enrollments inherent in this self-evolving inference process.
\item Experiments on various domains demonstrate that our method significantly enhances the generalization capability of TSE models, particularly in OOD scenarios, with minimal fine-tuning requirements.
\end{itemize}

\section{Related Work}
\textbf{Target Speaker Extraction:}
Current TSE research follows two approaches: embedding-based and embedding-free frameworks. The former utilizes a speaker encoder to extract identity-discriminative embeddings, either through pre-trained speaker verification models like TEA-PSE family~\cite{TEAPSE, TEAPSE2, TEAPSE3} or through jointly trained encoders as seen in X-TF-GridNet~\cite{HAO2024102550}.  In contrast, embedding-free frameworks, pioneered by the SpEx family~\cite{spex, spexplus, SpEx++}, perform joint modeling of mixture and enrollment in a shared latent space to avoid global embedding bottlenecks. Hybrid schemes have also been explored to combine global identity embeddings with raw spectral features to enhance robustness~\cite{bsrnn_tfmap, HeShulin, Shulin}. Building upon embedding-free frameworks, recent state-of-the-art architectures have introduced Attention-driven mechanisms for enhanced feature interaction. Notably, USEF-TSE~\cite{USEFTSE} employs a multi-head cross-attention mechanism to achieve early stage fusion of mixture and enrollment features. When coupled with powerful backbones such as TF-GridNet~\cite{tfgrid}, it significantly improves extraction quality. In the related Audio-Visual TSE domain, MoMuSE~\cite{muse,momuse} uses a memory bank to maintain a speaker identity momentum. Recent work such as TS-SEP~\cite{TSsep} has also explored joint diarization and separation to handle long-duration recordings by iteratively refining speaker embeddings. Despite their performance, these models typically operate in a fixed-enrollment inference setting, where the fixed enrollment signal fails to account for the identity drift.

\textbf{Retrieval-Augmented Generation in Audio Tasks:}
RAG was originally introduced in Natural Language Processing to enhance language models by integrating non-parametric memory~\cite{RAG}. This paradigm has recently expanded into the audio domain. In spoken dialogue systems, RAG has been utilized to facilitate end-to-end audio retrieval from hybrid knowledge bases~\cite{WavRAG}. For automated audio captioning, several works have leveraged retrieval mechanisms to fetch similar captions from external datastores to provide richer contextual descriptions~\cite{CAPRAG}. Furthermore, in text-to-audio generation, retrieval-augmented frameworks have been employed to mitigate data scarcity by using retrieved acoustic priors to guide the synthesis of rare events~\cite{TTARag1}. While effective for providing external priors in generation and captioning, RAG remains largely unexplored in TSE. Unlike existing frameworks relying on static databases, our EvoTSE introduces a self-evolving memory to dynamically track the target's identity drift through historical information.

\textbf{Mitigation of Speaker Confusion:}
Speaker confusion, the phenomenon where a model mistakenly tracks and extracts an interfering speaker, remains a major obstacle for reliable TSE deployment. Existing research addresses this challenge through three primary strategies. The first involves optimizing training objectives, where researchers integrate metric learning~\cite{TargetConfusion} or formulate chunk-level loss schemes~\cite{xsep_chunkloss} to penalize confusion errors and force the model to focus on fine-grained identity cues. The second strategy focuses on posterior verification and rectification. For instance, some frameworks utilize an auxiliary speaker verification module to detect inactive speaker cases~\cite{tse_sv} or implement dual-branch architectures that swap output streams upon detecting identity mismatches~\cite{refine}. The third strategy adopts data-driven augmentation, using resampling and rescaling pipelines to create diverse pseudo-speakers, thereby enhancing the model's discriminative generalization~\cite{TargetConfusionAug}. Unlike these methods that focus on penalizing errors or post-processing, our work uses retrieval-augmented enrollment to mitigate confusion at the input stage.

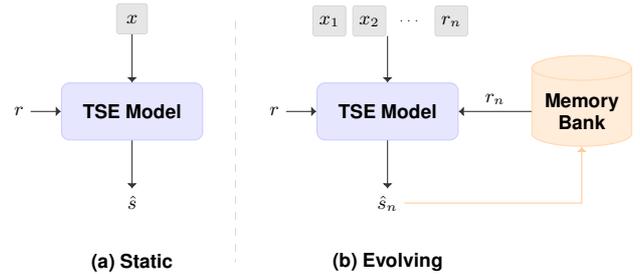
\begin{figure}[t]
\centering
\begin{tikzpicture}[
    scale=0.8, 
    every node/.style={transform shape, font=\sffamily}, 
    node distance=0.8cm,
    block/.style={rectangle, fill=blue!10, draw=blue!20,rounded corners=3pt, minimum width=2cm, minimum height=0.8cm, align=center, inner sep=10pt,font=\normalsize\sffamily\bfseries},
    seg_x/.style={rectangle, fill=gray!20, draw=gray!30, rounded corners=1pt, minimum width=0.5cm, minimum height=0.5cm, font=\sffamily\small},
    database/.style={cylinder, fill=orange!15, draw=orange!30, shape border rotate=90, aspect=0.25, minimum height=1.0cm, minimum width=1.5cm,align=center,inner sep=6pt,font=\normalsize\sffamily\bfseries},
    arrow/.style={-{Latex[length=2pt, width=4pt]}, thin, draw=black!80}
]

\node[block] (model_a) {TSE Model};
\node[seg_x] (X_long) [above=of model_a] {$x$};
\node (a_in) [left=0.5cm of model_a] {$r$};
\node (s_out) [below=of model_a] {$\hat{s}$};

\draw[arrow] (X_long) -- (model_a);
\draw[arrow] (a_in) -- (model_a);
\draw[arrow] (model_a) -- (s_out);

\node[font=\sffamily\bfseries] at (0,-2.5) {(a) Static};

\draw[dashed, gray!30] (1.7, 1.5) -- (1.7, -2.5);

\begin{scope}[xshift=4.2cm]
    \node[seg_x] (x2) at (-0.3, 1.5) {$x_2$};
    \coordinate (xv) at (0, 1.25) {};
    \node[seg_x] (x1) [left=0.1cm of x2] {$x_1$};
    \node (xdots) [right=0.1cm of x2, font=\tiny] {$\dots$};
    \node[seg_x] (xn) [right=0.1cm of xdots] {$r_n$};

    \node[block] (model_b) at (0, 0) {TSE Model};
    \node (a_b) [left=0.5cm of model_b] {$r$};
    \node[database] (M) [right=1.2cm of model_b] {Memory \\ Bank};
    \node (sn_out) [below=of model_b] {$\hat{s}_n$};

    \draw[arrow] (xv.south) -- (model_b.north); 
    \draw[arrow] (a_b) -- (model_b);
    \draw[arrow] (model_b) -- (sn_out);
    
    \draw[arrow] (M.west) -- (model_b.east) node[midway, above] {$r_n$};
    
    \draw[arrow, orange!40] (sn_out.east) -| (M.south) 
        node[pos=0.5, right, black] {};

    \node[font=\sffamily\bfseries] at (0,-2.5) {(b) Evolving};
\end{scope}

\end{tikzpicture}
\caption{System architecture: (a) Static enrollment. (b) Evolving enrollment with a right-side memory bank for state updates.}
\label{fig:TSE_Optimized}
\end{figure}
\section{Problem Formulation}
\subsection{Static TSE }
The objective of TSE is to isolate the target signal $s(t)$ from a multi-talker mixture $x(t)$, guided by a reference enrollment $r(t)$ of the target speaker. A generalized acoustic mixture in a reverberant environment is formulated as:
\begin{align}
x(t) = s(t) * h_s(t) + \sum_{i} n_i(t) * h_{n,i}(t) + \sum_{j} v_j(t) * h_{v,j}(t)
\end{align}
where $h(t)$ denotes the Room Impulse Response, while $n_i(t)$ and $v_j(t)$ represent additive $i$-th noise and $j$-th interfering speaker, respectively. In this study, to facilitate a focused analysis of identity ambiguity and speaker confusion which are the primary bottlenecks in challenging scenarios, we simplify the setup to an anechoic, noise-free environment with a single interferer:
\begin{align}
x(t) = s(t) + v(t)
\end{align}

Conventional TSE frameworks typically operate on short, isolated segments via a static mapping $\hat{s} = \text{Model}(x, r)$. This pipeline assumes that each segment is an independent sample relative to the static enrollment. Consequently, the model fails to accumulate speaker knowledge throughout a session, treating each inference as an isolated event with no temporal correlation.

\subsection{Evolving TSE}
In practical applications like voice assistants, the target speaker is persistent over extended durations, yet their acoustic characteristics such as emotional state and vocal effort frequently exhibit temporal drift. A static initial enrollment $r$ is often insufficient to cover these dynamic variations.

To bridge this gap, we reformulate the task into an evolving TSE pipeline, transforming the process from independent mapping into relational inference. Given a long-duration mixture $X = \{x_1, x_2, \dots, x_N\}$, the process is defined as:
\begin{align}
\hat{s}_n = \text{Model}(x_n, r, \mathcal{M}_{n-1}), \quad \text{for } n =1, 2, \dots, N
\end{align}
where $\mathcal{M}_{n-1}$ represents a state variable (memory bank) that summarizes the historical speaker cues distilled from the preceding $n-1$ estimates. This framework allows the model to adapt to OOD variations (e.g., emotional shifts) by explicitly leveraging contextually relevant information from the ongoing session, thereby mitigating the root causes of speaker confusion. Similar to CSS~\cite{CSS} in segment-wise 
processing, our approach targets enrollment-guided single-speaker 
extraction instead of separating all sources.

\section{Proposed Method}
\label{sec:method}
\subsection{Framework Overview}
The EvoTSE framework redefines TSE as a retrieval-augmented task. EvoTSE transforms the conventional static mapping into an evolving, evidence-accumulating system. As illustrated in Fig. \ref{fig:overview}, for each incoming mixture segment $x_n$ in a long-duration session, EvoTSE operates through a closed-loop feedback pipeline designed to adapt to acoustic drift. The core philosophy of this approach is to treat the initial enrollment $r$ not as a fixed reference, but as a foundational reference to initiate a continuous knowledge discovery process. To achieve this, the architecture integrates several specialized components:
\begin{itemize}
\item \textbf{Contextual Retriever:} To bridge the domain gap, EvoTSE queries the memory bank to fetch historical estimates ($m_k$) that are most similar to the current mixture segment.
\item \textbf{Backbone Extractor:} A backbone TSE network that leverages the enhanced enrollment signal ($\text{r}_n$) to isolate the target speaker's voice from the interfering speakers.
\item \textbf{Reliability Classifier:} A gated admission module that validates the identity consistency of each estimate to prevent memory poisoning and error propagation.
\item \textbf{Memory Curator:} A memory management module that employs a replacement policy to maximize the acoustic diversity of the memory bank within a fixed capacity.
\end{itemize}

For a specific segment $x_n$, as illustrated in Fig. \ref{fig:overview}, the contextual retriever utilizes the mixture itself as a query to retrieve the Top-$k$ relevant entries from the memory bank ($\mathcal{M}_{n-1}$). EvoTSE then fuses the initial enrollment $r$ with these retrieved entries to generate an acoustically matched signal $\text{r}_n$. Using this expanded enrollment, the extractor produces the clean estimate $\hat{s}_n$. To ensure the stability of the evolution, $\hat{s}_n$ is passed to the reliability classifier for an identity-consistency check. Only verified estimates are forwarded to the Evolution module to update the memory bank, thereby preparing EvoTSE with refined speaker knowledge for the subsequent segment $x_{n+1}$.

\usetikzlibrary{arrows.meta, positioning, calc, shapes.geometric, backgrounds, fit}

\begin{figure*}[t]
  
  \centering
  
  \begin{subfigure}{\textwidth}
    
    \centering
    \begin{tikzpicture}[
        scale=0.85, every node/.style={transform shape},
        node distance=2.0cm and 1.6cm,
        font=\sffamily\scriptsize,
        module/.style={rectangle, rounded corners=3pt, minimum width=2.2cm, minimum height=1cm, align=center, draw=blue!20, fill=blue!10,inner sep=8pt,font=\normalsize\sffamily\bfseries},
        storage/.style={cylinder, shape border rotate=90, minimum width=1.5cm, minimum height=1.6cm, aspect=0.2, align=center, fill=orange!15,draw=orange!30,inner sep=8pt,font=\normalsize\sffamily\bfseries},
        data/.style={rectangle, draw=none, minimum width=2.0cm, minimum height=1.0cm, fill=none, align=center},
        data_mix/.style={rectangle, draw=none, minimum width=2.0cm, minimum height=1.5cm, fill=none, align=center},
        gate/.style={rectangle, rounded corners=3pt, minimum width=2.2cm, minimum height=1cm, align=center, draw=orange!30, fill=orange!15,inner sep=8pt,font=\normalsize\sffamily\bfseries},
        arrow/.style={-{Latex[length=2pt, width=4pt]}, thin, draw=black!80},
        concat/.style={circle, draw=black!80, fill=white!10, inner sep=2pt, font=\small\sffamily\bfseries, minimum size=0.3cm}
      ]
        \coordinate (seq_base) at (-3.5, 0);
        \node[data_mix] (m1) at ($(seq_base) + (0, 1.5)$) {};
        \node[left=0.1cm of m1, font=\bfseries] {$x_1$};
        \drawwaveform{10}{blue!80}{m1.center}{0.6} \drawwaveform{44}{red!80}{$(m1.center)+(0.02,0)$}{0.6}
        
        \node[data_mix] (mn) at  (seq_base) {};
        \node[left=0.1cm of mn, font=\bfseries] {$x_n$};
        \drawwaveform{1}{blue!80}{mn.center}{0.6} \drawwaveform{99}{red!80}{$(mn.center)+(0.02,0)$}{0.6}
        
        \node at ($(seq_base) + (0, 0.65)$) {\vdots};
        
        \node[data_mix] (mm) at ($(seq_base) + (0, -1.5)$) {};
        \node[left=0.1cm of mm, font=\bfseries] {$x_m$};
        \drawwaveform{20}{blue!80}{mm.center}{0.6} \drawwaveform{77}{red!80}{$(mm.center)+(0.02,0)$}{0.6}
        \node at ($(seq_base) + (0, -0.65)$) {\vdots};
    
        \node[data_mix] (mix) at (0, 0) {};
        \node[above=-10pt of mix] {Mixture $x_n$};
        \drawwaveform{1}{blue!80}{mix.center}{0.6}
        \drawwaveform{99}{red!80}{$(mix.center)+(0.02,0)$}{0.6}
        \draw[arrow, dashed, gray!60] (mn) -- (mix);
    
        \node[module] (retriever) [right=of mix] {Contextual \\ Retriever};
        \node[concat] (builder) [right=2.0cm of retriever] {C};
        \node[module] (extractor) [right=2.0cm of builder] {Extractor $F_\theta$};
        \node[data] (est) [right=0.8cm of extractor] {};
        \node[above=-2pt of est] {Estimate $\hat{s}_n$};
        \drawwaveform{1}{blue!80}{est.center}{0.6}
        
        \node[data] (pioneer) [above=1.8cm of builder] {};
        \node[above=-2pt of pioneer] {Init Enroll $r$};
        \drawwaveform{5}{blue!80}{pioneer.center}{0.6}
    
        \coordinate (top_padding) at ($(mix.north) + (0, 0.85)$);
    
        \node[storage] (memory) at ($(retriever) + (0, -3.5)$) {Memory \\ Bank $\mathcal{M}_{n-1}$};
        \node[module, draw=orange!30, fill=orange!15] (evictor) at ($(builder) + (0, -3.5)$) {Memory \\ Evolution};
        \node[gate] (classifier) at ($(extractor) + (0, -3.5)$) {Classifier};
    
        \begin{scope}[on background layer]
            \node [draw=blue!20, dashed, fill=blue!2, inner sep=12pt, 
                   fit=(mix) (retriever) (builder) (extractor) (top_padding)] (proc_group) {};
            \node [draw=orange!20, dashed, fill=orange!2, inner sep=12pt, fit=(memory) (evictor) (classifier)] (mem_group) {};
        \end{scope}
    
        \draw [arrow] (mix) -- node[midway, above] {Query} (retriever);
            \draw [arrow] (retriever) -- node[midway, above] {k$\times m$} (builder);
        
        \coordinate (bypass_y) at ($(mix.north) + (0, 0.5)$); 
        \draw [arrow, rounded corners=6pt] 
            (mix.north) -- 
            (mix.north |- bypass_y) -- 
            (extractor.north |- bypass_y) -- 
            (extractor.north);
    
        \def\jumpsize{0.15} 
        \draw [arrow,-] (pioneer.south) -- ($(builder.north |- bypass_y) + (0, \jumpsize)$);
        \draw [arrow,-] ($(builder.north |- bypass_y) + (0, \jumpsize)$) 
              arc (90:-90:\jumpsize); 
        \draw [arrow] ($(builder.north |- bypass_y) + (0, -\jumpsize)$) -- (builder.north);
    
        \draw [arrow] (builder) -- node[midway, above, font=\footnotesize] {$r_n$} (extractor);
        \draw [arrow] (extractor) -- (est);
    
            \draw [arrow] (memory) -- node[midway, right] {Retrieve} (retriever);
        \draw [arrow, rounded corners=8pt] (est.south) -- ($(est.south |- classifier.east)$) -- (classifier.east);
        \draw [arrow] (classifier) -- node[midway, above] {Accept} (evictor);
        \draw [arrow] (evictor) -- node[midway, above] {Evict} (memory);
        
        \draw [dashed, gray, -{Latex[length=2pt, width=4pt]}, rounded corners=10pt, thick] 
            (mem_group.west) -- ($(mix.south |- mem_group.west)$) -- (proc_group.south -| mix.south) 
            node[pos=0.6, left, black, xshift=-2pt] {$n \to n+1$};
    
        \node[anchor=south west, font=\scriptsize\itshape, color=blue!50] at (proc_group.south west) {Retrieval};
        \node[anchor=south west, font=\scriptsize\itshape, color=orange!80] at (mem_group.south west) {Evolution};
    
    \end{tikzpicture}

    \caption{Overview}
    \label{fig:overview}
  \end{subfigure}

  \vspace{0.2cm}

  \begin{subfigure}{0.48\textwidth}
    \centering
    \begin{tikzpicture}[
            scale=0.82, transform shape,
            font=\sffamily\scriptsize,
            >=Stealth,
            module/.style={rectangle, rounded corners=3pt, minimum width=2.2cm, minimum height=1cm, align=center, draw=blue!20, fill=blue!10,inner sep=8pt,font=\normalsize\sffamily\bfseries},
            storage/.style={cylinder, shape border rotate=90, minimum width=1.5cm, minimum height=1.6cm, aspect=0.2, align=center, fill=orange!15,draw=orange!30,inner sep=8pt,font=\normalsize\sffamily\bfseries},
            vector/.style={rectangle, draw=red!30, fill=red!10, minimum width=0.25cm, minimum height=1.1cm},
            container/.style={draw=blue!20,fill=blue!2, dashed, rounded corners=2pt},
            arrow/.style={-{Latex[length=2pt, width=4pt]}, thin, draw=black!80,rounded corners=4pt}
        ]
        
            \coordinate (input_mix) at (0, 0);
            \coordinate (input_mix_below) at (0, -0.3);
            \coordinate (input_mix_left) at ($(input_mix)+(-0.4,0)$) ;
            
            \drawwaveformshort{24}{blue!60}{input_mix}{1.2}
            \drawwaveformshort{77}{red!50}{input_mix}{0.9}
            \node[above=0.3cm of input_mix] {Mix};
            
            \node[module] (encoder) at (2.4, 0) {ECAPA\\\&Emo2Vec};
            \draw[arrow] ($(input_mix)+(0.6,0)$) -- (encoder);
            
            \node[vector] (query_vec) [right=0.5cm of encoder] {};
            \node[above=0.05cm of query_vec] {Emb};
            \draw[arrow] (encoder) -- (query_vec);
            
            \node[vector] (v1) at ($(query_vec)+(1.2, 0.45)$) {};  
            \node[vector] (v2) at ($(v1)+(0.6, -0.45)$) {};       
            \node[vector] (v3) at ($(v2)+(0.6, -0.45)$) {};       
            
            \draw[arrow] (query_vec.east) -- +(0.25,0) |- (v1.west);
            \draw[arrow] (query_vec.east) -- (v2.west); 
            \draw[arrow] (query_vec.east) -- +(0.25,0) |- (v3.west);
            
            \node[draw=gray!40, fill=gray!5, fill opacity=0.1, inner sep=5pt, fit=(v1)(v3)] (pool_box) {};
            \node[storage] (database) at ($(pool_box.east)+(1.7, 0)$) {Memory \\ Bank};
            
            \coordinate (database_right) at ($(database)+(0.75,0)$) ;
            
            \draw[arrow] (pool_box.east) -- node[above] {TopK} (database);
        
            \node[module] (tse) at (2.4, -3.2) {Extractor};
            
            \draw[arrow,rounded corners=6pt](input_mix_below.south) -- +(0,-1.0) -- ($(tse.north)+(0, 0.8)$) -- (tse.north);

            \coordinate (output_est) at (0, -3.2);
            \coordinate (output_est_left) at ($(output_est)+(-0.4,0)$) ;
            \drawwaveformshort{24}{blue!60}{output_est}{1.3}
            \node[above=0.3cm of output_est] {Est};
            \draw[arrow] ($(output_est)+(0.6,0)$) -- (tse.west);
        
            \node[draw=none, inner sep=0pt] (aux_frame) at (database.center |- tse.center) {
                \begin{tikzpicture}
                    \coordinate (r1) at (0, 0.8); \drawwaveformshort{15}{blue!50}{r1}{0.7}

                    \coordinate (r2) at (0, 0.2); \drawwaveformshort{16}{blue!50}{r2}{0.7}
                    \node at (0, 0.6) {\vdots};
                    \node[below=0.3cm of r2] {Retrieved};
                    
                    \draw[dashed, gray!40] (-0.75, -0.35) -- (0.75, -0.35);
                    
                    \coordinate (ia) at (0, -0.7); \drawwaveformshort{17}{blue!50}{ia}{0.7}
                    \node[below=0.3cm of ia] {Init Enroll};
                \end{tikzpicture}
            };

            \draw[arrow] (database.south) -- (aux_frame.north);
        
            \coordinate (merged_center) at ($(tse.east)!0.5!(aux_frame.west)$);
            \path ($(merged_center)-(0.85,0)$) coordinate (p1);
            \path (merged_center)             coordinate (p2);
            \path ($(merged_center)+(0.85,0)$) coordinate (p3);
            
            \drawwaveformshort{15}{blue!50}{p1}{0.7}
            \drawwaveformshort{16}{blue!50}{p2}{0.7}
            \drawwaveformshort{17}{blue!50}{p3}{0.7}
            
            \draw[arrow] (aux_frame.west) -- ($(p3)+(0.45,0)$);
            \draw[arrow] ($(p1)-(0.45,0)$) -- (tse.east);
            \node[above=0.3cm of p2] {Enroll};
            
            \begin{scope}[on background layer]
                \node[container, fit=(input_mix)(encoder)(database)(v1)(pool_box)(input_mix_left)(database_right), inner sep=6pt] (top_zone) {};
                
                \node[container, fit=(tse)(output_est)(aux_frame)(output_est_left), inner sep=6pt] (bottom_zone) {};
                
            \end{scope}
        
        \end{tikzpicture}
    \caption{Retrieval}
  \label{fig:Retrieval}
  \end{subfigure}
  \hfill
  \begin{subfigure}{0.48\textwidth}
    \centering
    \begin{tikzpicture}[
            scale=0.82, transform shape,
            font=\sffamily\fontsize{6.5pt}{7pt}\selectfont,
            >=Stealth,
            module/.style={rectangle, rounded corners=3pt, minimum width=2.2cm, minimum height=1cm, align=center, draw=blue!20, fill=blue!10,inner sep=8pt,font=\normalsize\sffamily\bfseries},
            vector/.style={rectangle, draw=red!30, fill=red!10, minimum width=0.25cm, minimum height=1.3cm},
            uniformbox/.style={rectangle, draw=blue!15, fill=blue!5, minimum width=0.5cm, minimum height=0.4cm, align=center,rounded corners=2pt},
            dashedbox/.style={rectangle, draw=blue!25, fill=blue!1, minimum width=0.5cm, minimum height=0.4cm, align=center,rounded corners=2pt,dashed},
            container/.style={draw=orange!30,fill=orange!2, dashed, rounded corners=2pt},
            arrow/.style={-{Latex[length=2pt, width=4pt]}, thin, draw=black!80}
        ]
        
            \coordinate (input_mix) at (0, -1.3);
            \coordinate (input_mix_left) at (-1, 0);
            \drawwaveformshort{1}{blue!60}{input_mix}{1.2}
            \node[left=0.5cm of input_mix,font=\sffamily\scriptsize] {Est};
            
            \node[module] (encoder) at (0, 0) {ECAPA};
            \draw[arrow] (0.0,-0.9) -- (encoder);
            
            \node[vector] (query) at ($(encoder)+(1.8, 0.0)$) {};
            \node[above=0.05cm of query,font=\sffamily\scriptsize] {Emb};
            \draw[arrow] (encoder) -- (query);
            
            \node[vector] (v1) at ($(query)+(1.2, 0.45)$) {};
            \node[vector] (v2) at ($(v1)+(0.7, -0.45)$) {};       
            \node[vector] (v3) at ($(v2)+(0.7, -0.45)$) {};    
         
            
            \draw[arrow,rounded corners=4pt] (query.east) -- (v2.west);
            \draw[arrow,rounded corners=4pt] (query.east) -- +(0.4, 0) -- +(0.4, 0.45) -- (v1.west);
            \draw[arrow,rounded corners=4pt] (query.east) -- +(0.4, 0) -- +(0.4, -0.45) -- (v3.west);
            
            \node[draw=gray!40, fill=gray!5, fill opacity=0.1, inner sep=5pt, fit=(v1)(v3)] (pool_box) {};
        
            \node[uniformbox] (s1) at ($(v1)+(2.5, 0.0)$) {0.47}; 
            \node[uniformbox] (s2) at ($(s1)+(0.0, -0.45)$) {0.32};
            \node[uniformbox] (s3) at ($(s2)+(0.0, -0.45)$) {0.51};
            
            \draw[arrow] (v1.east) -- (s1);
            \draw[arrow] (v2.east) -- (s2);
            \draw[arrow] (v3.east) -- (s3);
        
            \node[circle,draw=green!40!black, thick, minimum size=0.5cm] (replace_icon) at ($(query)+(0.25,-3.2)$$) {};
          
            \node[green!40!black, below=0.05cm of replace_icon, font=\sffamily\scriptsize] {Replace};
        
            \coordinate (list_y) at (0, -3.2);
            
            \node[uniformbox] (n2) at ($(encoder)+(0.0, -3.2)$){0.57};
            \node[uniformbox] (n1) at ($(n2)+(-0.7, 0.0)$) {0.34};
            \node[uniformbox] (n3) at ($(n2)+(0.7, 0.0)$) {0.67}; 
            \coordinate (n1_left) at (-1, -3.8);
            
            \node[uniformbox, fill=blue!20] (r_v1) at (v1.center |- list_y) {0.75};
            \node[uniformbox] (r_v2) at (v2.center |- list_y) {0.57};
            \node[uniformbox] (r_v3) at (v3.center |- list_y) {0.67};
            
            \node[uniformbox] (avg_res) at (s3.center |- list_y) {0.34};
            \node at (avg_res.north) [above right=0.05cm,font=\sffamily\scriptsize] {avg};
        
            \draw[arrow,dashed,draw=green!40!black,thick,-] (query.south) -- +(0,-1.5) |- (replace_icon.west);
            \draw[->, arrow,dashed,draw=green!40!black,thick] (replace_icon.east) -- (v1.south);
            \draw[arrow] (replace_icon.west) -- (n3.east) ;
            \draw[arrow] (r_v1.west) -- (replace_icon.east) ;
        
            \draw[arrow,dashed] (s3.south) -- (avg_res.north);
            \draw[arrow] (avg_res.west) -- (r_v3.east);
        
            \node[dashedbox,font=\sffamily\scriptsize] (alpha_high) at ($(s1)+(1.3, 0.0)$){$\tau$=0.6};
            \node[dashedbox,font=\sffamily\scriptsize] (reject) at ($(alpha_high)+(1.2, 0.0)$) {Reject};
            \node[uniformbox,font=\sffamily\scriptsize] (alpha_low) at ($(s3)+(1.3, 0.0)$) {$\tau$=0.5};
            \node[uniformbox,font=\sffamily\scriptsize] (pass) at ($(alpha_low)+(1.2, 0.0)$) {Accept};
        
            \draw[arrow, dashed] (s3.east) -- +(0.2,0) |- (alpha_high.west);
            \draw[arrow, dashed] (alpha_high) -- (reject);
            \draw[arrow] (s3.east) -- +(0.3,0) |- (alpha_low.west);
            \draw[arrow] (alpha_low) -- (pass);
            \draw[arrow, rounded corners=4pt] (pass.south) |- (avg_res.east);
            \coordinate (top_limit) at (0, -2.6);
            \begin{scope}[on background layer]
                \node[container, fit=(input_mix)(pool_box)(reject)(pass)(input_mix_left), inner sep=10pt] (top_zone) {};
                \node[container, fit=(n1)(avg_res)(n1_left)(top_limit), inner sep=10pt] (bottom_zone) {};
            \end{scope}
        
        \end{tikzpicture}
    \caption{Evolution}
  \label{fig:Classification}
  \end{subfigure}
  
  \caption{(a) Overview: The overall architecture where speaker extraction is enhanced by historical cues. (b) Retrieval: The process of querying the memory bank using mixture embeddings to construct an acoustically matched enrollment. (c) Evolution: The reliability-gated evolution, where new estimates are validated via a threshold $\tau$ and integrated into the memory bank}
  \label{fig:main}
\end{figure*}
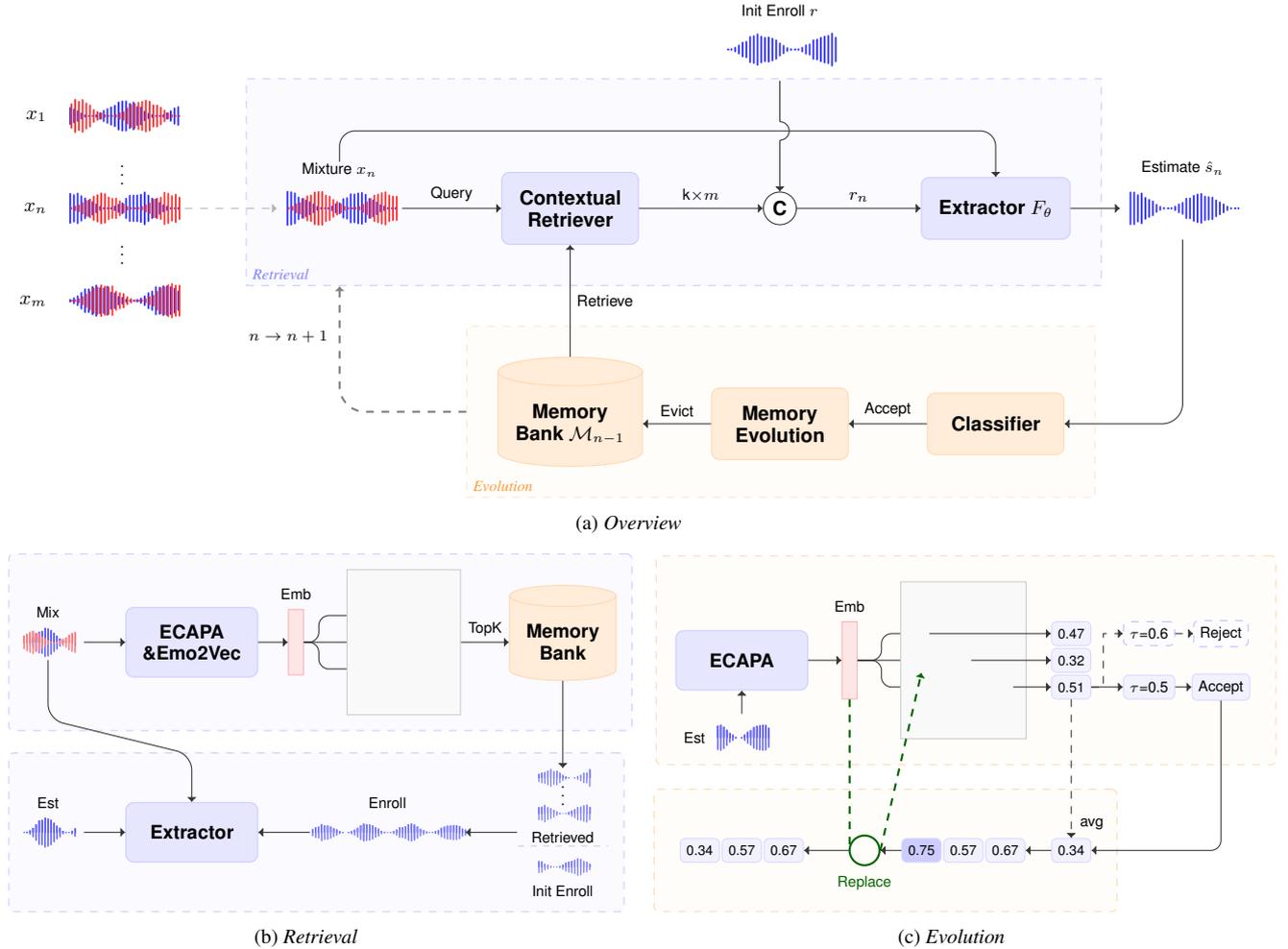

\subsection{Contextual Retriever}

The contextual retriever bridges the gap between the static initial enrollment and the time-varying mixture environment. We define the memory bank at step $n$ as $\mathcal{M}_{n-1} = \{m_1, m_2, \dots, m_{|\mathcal{M}|}\}$. As illustrated in top panel of Fig. \ref{fig:Retrieval}, the system employs a dual-stream independent retrieval pipeline using two pretrained encoders: ECAPA-TDNN~\cite{ecapa,wang2023wespeaker} for speaker identity $E_{\text{spk}}(\cdot)$ and Emotion2vec~\cite{ma2023emotion2vec} for emotional states $E_{\text{emo}}(\cdot)$. For each stream, the probability of an entry $m$ being relevant to the current mixture $x_n$ is formulated via its own similarity-based distribution:

\begin{align}
p_{\text{attr}}(m|x_n) = \frac{\exp \left( \text{sim}(E_{\text{attr}}(x_n), E_{\text{attr}}(m)) \right)}{\sum_{m_j \in \mathcal{M}_{n-1}} \exp \left( \text{sim}(E_{\text{attr}}(x_n), E_{\text{attr}}(m_j)) \right)}
\end{align}
where $\text{attr} \in \{\text{spk, emo}\}$, $\text{sim}(\cdot, \cdot)$ denotes cosine similarity. Specifically, the retrieval process consists of two parallel steps: Selecting the Top-$k$ segments $\mathcal{M}_{\text{spk}}$ with the highest speaker similarity, and selecting the Top-$k$ segments $\mathcal{M}_{\text{emo}}$ with the highest emotional similarity. The final contextual subset $\mathcal{M}_k$ is formed by the union of these two sets:
\begin{align}
\mathcal{M}_k = \mathcal{M}_{\text{spk}} \cup \mathcal{M}_{\text{emo}}
\end{align}
Consequently, the actual number of retrieved segments varies between $k$ and $2k$. For simplicity in subsequent discussions, we continue to denote the size of this set as $k$. 

\subsection{Backbone Extractor}
\label{sec:RetrievalQuantity}
During the extraction stage, EvoTSE performs identity-anchored recomposition to transform the retrieved discrete segments into a unified guidance signal. Let $\mathcal{M}_k = \{m_1, m_2, \dots, m_k\}$ represent the set of Top-$k$ retrieved segments. As illustrated in the bottom panel of Fig. \ref{fig:Retrieval}, the extended enrollment signal $r_n$ is constructed by temporally concatenating the initial enrollment $r$ with the $k$ retrieved historical segments:

\begin{align}
\text{r}_n = \text{Concat}(r, m_1, m_2, \dots, m_k)
\end{align}

In this sequence, the initial enrollment $r$ serves as a stable identity anchor to prevent speaker confusion, while the retrieved segments capture current acoustic variations. Subsequently, the backbone TSE extractor $F_\theta$ processes the mixture $x_n$ conditioned on this recomposed enrollment to produce the final estimate:
\begin{align}
\hat{s}_n = F_\theta(x_n, \text{r}_n)
\end{align}
The backbone model effectively mitigates the risk of speaker confusion by dynamically focusing on the enrollments that most closely match the current acoustic state.

\subsection{Reliability Classifier}
\label{sec:Classifier}
A critical challenge in evolving TSE is the risk of error propagation. If an interfering speaker is erroneously extracted and stored, the memory bank becomes poisoned, which leads to a cumulative failure in subsequent segments. To mitigate this while maximizing the acoustic diversity of the memory bank, we introduce a reliability classifier. For each new estimate $\hat{s}_n$, the classifier evaluates its identity validity by performing a nearest-neighbor check against the entire current memory bank $\mathcal{M}_{n-1}$. Using the pre-trained ECAPA-TDNN as the speaker embedding extractor $E_{\text{spk}}(\cdot)$, the reliability score $c_n$ is defined as the maximum cosine similarity between the current estimate and any existing entry in the memory bank:

\begin{align}
c_n = \max_{m \in \mathcal{M}_{n-1}} \left( \text{sim}(E_{\text{spk}}(\hat{s}_n), E_{\text{spk}}(m)) \right)
\end{align}
The admission logic is governed by a threshold $\tau$, which determines whether the memory bank $\mathcal{M}$ is updated as follows:
\begin{align}
\mathcal{M}_n = \begin{cases} \mathcal{M}_{n-1} \cup \{\hat{s}_n\}, & \text{if } c_n > \tau \\ \mathcal{M}_{n-1}, & \text{otherwise} \end{cases}
\end{align}

As illustrated in the top panel of Fig. \ref{fig:Classification}, the incoming estimate $\hat{s}_n$ generates a query embedding (e.g., Emb). This embedding is compared against all candidate vectors currently stored in the memory bank to compute individual similarity scores (e.g., 0.47, 0.32, 0.51) and identifies the maximum value (e.g., 0.51). Based on the predefined threshold $\tau$, a binary decision is made. If the peak similarity is below the threshold (e.g., $\tau$=0.6), the segment is labeled as a rejection. Conversely, if the score meets the criteria (e.g., $\tau$=0.5), the segment is accepted and is prepared for memory integration. This mechanism ensures that only segments with a high degree of identity confidence are allowed to influence future extraction steps.

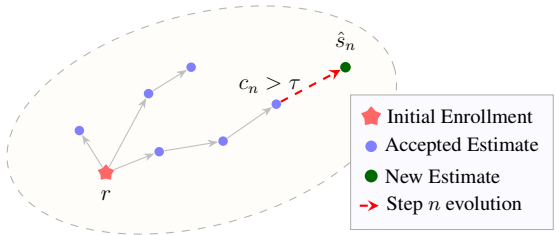
\begin{figure}[t]
\centering
\begin{tikzpicture}[
    scale=0.7, transform shape,
    node distance=1.5cm,
    anchor_node/.style={star, star points=5, fill=red!60, inner sep=2.2pt},
    memory_node/.style={circle, fill=blue!50, inner sep=1.8pt},
    new_node/.style={circle, fill=green!40!black, inner sep=2pt},
    manifold/.style={ellipse, draw=gray!60, fill=orange!2, dashed, minimum width=7.5cm, minimum height=4cm, rotate=15}
]
    
    \node[manifold] (m) at (1.8, 1.0) {};
    
    \node[anchor_node] (A) at (0,0) {}; 
    \node[below=0.1cm of A,font=\large] {$r$};
    
    \node[memory_node] (n1) at (1.0, 0.4) {};
    \node[memory_node] (n2) at (2.2, 0.6) {};
    \node[memory_node] (n3) at (3.2, 1.3) {};
    
    \node[memory_node] (n4) at (0.8, 1.5) {};
    \node[memory_node] (n5) at (1.6, 2.0) {};
    \node[memory_node] (n6) at (-0.5, 0.8) {};
    
    \node[new_node] (Target) at (4.5, 2.0) {};
    \node[above=0.1cm of Target, font=\large] {$\hat{s}_n$};
    
    \begin{scope}[every edge/.style={draw, -{Stealth[scale=0.8]}, gray!50, thin}]
        \path (A) edge (n1);
        \path (n1) edge (n2);
        \path (n2) edge (n3);
        \draw[->, >=stealth, thick, red, dashed] (n3) to node[midway, left, black, xshift=-2pt,yshift=0.4pt,font=\large] {$c_n > \tau$} (Target);
        
        \path (A) edge (n4);
        \path (n4) edge (n5);
        \path (A) edge (n6);
    \end{scope}
    
    \matrix [draw, draw=gray!60, fill=blue!2, nodes={scale=0.85}] at (6.5,0.2) {
      \node [anchor_node, label=right:Initial Enrollment] {}; \\
      \node [memory_node, label=right:Accepted Estimate] {}; \\
      \node [new_node, label=right:New Estimate] {}; \\
      \draw [->, >=stealth, thick, red, dashed] (-0.15,0) -- (0.1,0) node[right, black] {Step $n$ evolution}; \\
    };
    
\end{tikzpicture}
\caption{Conceptual illustration of speaker identity evolution on the manifold. }
\label{fig:step_evolution}
\end{figure}

The reliability score $c_n$ is calculated by comparing a new estimate with every segment already in the memory bank, rather than just the initial enrollment. This creates a "bridge" effect in the speaker's feature space. As shown in Fig. \ref{fig:step_evolution}, if the initial recording is highly emotional, a neutral segment might be too different to be accepted directly. However, the system can gradually bridge this gap by first admitting moderately emotional segments that share similarities with both sides. By using these intermediate segments as stepping stones, the memory bank effectively expands its coverage to include a wide range of vocal styles while still ensuring that every new addition is verified as the correct speaker.

\subsection{Memory Curator}
To maintain computational efficiency and prevent the memory bank from being overwhelmed by repetitive acoustic information, we constrain its capacity to a maximum of $|\mathcal{M}|_{\max}$ entries. When $\mathcal{M}$ reaches this limit, a redundancy-aware eviction policy is triggered. For each entry $m_i \in \mathcal{M}$, we calculate a global redundancy score $\Omega_i$, which quantifies its average similarity to all other samples in the bank. EvoTSE identifies and removes the entry with the highest $\Omega_i$ score. This ensures that the memory bank contains a wide range of the target speaker’s acoustic characteristics rather than redundant variations of the same state. The $\Omega_i$ score is defined as follows, where $\alpha$ is a weighting factor used to balance the influence between emotional similarity and speaker similarity:

\begin{equation}
\begin{split}
\Omega_i = \frac{1}{|\mathcal{M}|_{\max} - 1} \sum_{j \neq i} \big( &\text{sim}(E_{\text{spk}}(m_i), E_{\text{spk}}(m_j)) + \\
\alpha \cdot &\text{sim}(E_{\text{emo}}(m_i), E_{\text{emo}}(m_j)) \big)
\end{split}
\end{equation}

Unlike the reliability classifier, memory curator also utilizes Emotion2vec. However, to maintain the clarity of the illustration, this component is not explicitly shown. As illustrated in the bottom panel of Fig. \ref{fig:Classification}, once an estimate passes the reliability check, its redundancy score (e.g., 0.34) is compared with existing entries (e.g., 0.34, 0.57, 0.67). The entry exhibiting the highest redundancy (e.g., 0.75) is targeted for eviction. The new estimate then replaces this redundant entry, which effectively enriches the memory bank's diversity.

\subsection{Training Strategy: Artifact-aware Learning}
We propose an Artifact-aware Learning strategy, which is divided into two progressive stages and shifts from static feature mapping to evolving identity alignment.

In the initial stage, the TSE extractor is trained in a conventional static pipeline. The goal is to establish a primary mapping between the mixture $x$ and the target speaker $s$, guided by enrollment $r$. This stage provides the model with the basic capability to isolate the target speaker, which significantly accelerates the convergence of the subsequent sequential training.

Building upon this foundation, the second stage introduces \emph{alignment fine-tuning} where the first stage extractor is trained on sequences rather than isolated pairs. Unlike traditional static methods where one enrollment $r$ corresponds to a single mixture $x$, this stage adopts group-based training. Each training sample consists of an initial enrollment $r$, a group of $N$ consecutive mixtures $\{x_1, \dots, x_N\}$, and their corresponding ground truth targets $\{s_1, \dots, s_N\}$. Within each group, the model processes mixtures sequentially. For the $n$-th mixture $x_n$, the enrollment $\text{r}_n$ is refined by the EvoTSE module using information captured from the previous $n-1$ estimates. This process is implemented with mixtures and targets shaped as $(B, N, T)$ and initial enrollments as $(B, T)$, where the group serves as the minimum unit for gradient calculation and backpropagation as follows:
\begin{align}
\mathcal{L}_{total} = \frac{1}{B \cdot N} \sum_{b=1}^B \sum_{n=1}^N \text{Loss}(\hat{s}_{bn}, s_{bn})
\end{align}

This strategy ensures training-inference consistency by adapting the TSE backbone to the evolving enrollment format of EvoTSE. Secondly, it enhances artifact robustness. Neural extractors naturally introduce subtle signal distortions, which leads to significant degradation. By backpropagating through the entire chain, the model learns to extract invariant speaker identity from distorted references.

\section{Experimental Setup}
\subsection{Datasets}
Following the configuration of the backbone models, we use the WSJ0-2mix dataset~\cite{dc} for fundamental training and evaluation. It consists of three subsets: the training set with 20,000 utterances from 101 speakers, the development set with 5,000 utterances from 101 speakers, and the test set with 3,000 utterances from 18 speakers. For brevity, we refer to this training setup as WSJ in subsequent sections. This dataset is derived from the Wall Street Journal (WSJ0) corpus~\cite{garofolo1993csr}. Furthermore, we incorporate the max version of Libri2mix-clean~\cite{Cosentino2020LibriMixAO} test set, consisting of 3,000 mixtures from 40 non-overlapping speakers. For each mixture, we exclusively use Source 1 as the target speaker for extraction, resulting in 3,000 evaluation samples. While WSJ0-2mix and Libri2mix-clean is a standard benchmark, its acoustic conditions are relatively stationary and cannot reflect the speaker confusion caused by emotional shifts. To address this, we introduce the Emotional Speech Database (ESD)~\cite{zhou2022emotional,zhou2021seen} to evaluate model robustness in more challenging scenarios. We construct an ESD-based training and test set. The test set contains 4,986 mixtures from 4 speakers (2 males and 2 females), covering various emotional states including Angry, Happy, Neutral, Sad, and Surprise. The remaining data is processed following the same construction method as WSJ0-2mix to form the ESD training set. This setup provides a benchmark for tracking speaker identity under significant acoustic variations. For evaluation under both EvoTSE and Static protocols, test mixtures across all datasets are grouped by target speaker identity, with a single initial enrollment randomly selected per group. To maintain consistency with the backbone USEF-TFGridNet~\cite{USEFTSE}, all datasets are resampled to 8 kHz.

\subsection{Evaluation Metrics}
Evaluation Metrics: We employ three key metrics to quantify performance:

\textbf{SI-SDRi (dB):} The scale-invariant signal-to-distortion ratio improvement of the estimated target $\hat{s}_n$ relative to the mixture $x_n$.We use the SI-SDRi to evaluate the overall quality.

\textbf{NSR (\%):} Following the definition in X-TaSNet~\cite{xtas}, we adopt the Negative SI-SDRi Rate (NSR) as an objective metric to quantify speaker confusion. NSR measures the proportion of extracted segments that yield a negative SI-SDRi. As observed in~\cite{xtas}, a significant negative SI-SDRi typically occurs when the model erroneously locks onto an interfering speaker. Thus, NSR serves as a reliable approximation for the speaker extraction error rate.

\textbf{SI-SDRiC (dB):} Similar to the conditional evaluation approach used in~\cite{refine}, we introduce SI-SDRi of Correct samples (SI-SDRiC) to decouple the quality of signal reconstruction from the frequency of identity confusion. Specifically, SI-SDRiC is defined as the average SI-SDRi calculated exclusively over the subset of segments where no speaker confusion occurs (i.e., the $1 - \text{NSR}$ portion of the data). This metric shows the potential performance of the extraction back-end when the speaker is correctly identified.

\begin{table*}[t]
\centering
\footnotesize
\caption{Performance comparison on WSJ0-2mix, ESD-test, and Libri2mix-clean.}
\label{tab:main_results}
\setlength{\tabcolsep}{5.5pt} 
\begin{tabular}{@{} l l ccc ccc ccc @{}} 
\toprule
\textbf{Train} & \textbf{Method} & \multicolumn{3}{c}{\textbf{WSJ0-2mix}} & \multicolumn{3}{c}{\textbf{ESD-test}} & \multicolumn{3}{c}{\textbf{Libri2mix-clean}} \\ 
\cmidrule(lr){3-5} \cmidrule(lr){6-8} \cmidrule(lr){9-11}
& & \makecell{SI-SDRi\\(dB)$\uparrow$} & \makecell{NSR\\(\%)$\downarrow$} & \makecell{SI-SDRiC\\(dB)$\uparrow$} & \makecell{SI-SDRi\\(dB)$\uparrow$} & \makecell{NSR\\(\%)$\downarrow$} & \makecell{SI-SDRiC\\(dB)$\uparrow$} & \makecell{SI-SDRi\\(dB)$\uparrow$} & \makecell{NSR\\(\%)$\downarrow$} & \makecell{SI-SDRiC\\(dB)$\uparrow$} \\ \midrule

\multirow{4}{*}{WSJ} 
 & USEF-TFGridNet (Standard) & 23.35 & 0.6 & \textbf{23.59} & 1.81 & 24.3 & 13.28 & 16.51 & 4.7 & 18.41 \\ 
 & USEF-TFGridNet (Static)   & 23.01 & 1.2 & 23.44 & 2.09 & 23.9 & 13.37 & 16.65 & 4.6 & 18.53 \\
 & EvoTSE ($k$=3)        & \underline{23.40} & \underline{0.5} & \underline{23.58} & \underline{10.73} & \underline{8.1} & \underline{13.45} & \underline
{17.91} & \underline
{2.2} & \underline
{18.71} \\ 
 & EvoTSE ($k$=24)       & \textbf{23.44} & \textbf{0.4} & \underline{23.58} & \textbf{11.34} & \textbf{6.4} & \textbf{13.59} & \textbf{17.92} & \textbf{2.1} & \textbf{18.72} \\ \midrule

\multirow{4}{*}{\makecell[l]{WSJ + \\ ESD}} 
 & USEF-TFGridNet (Standard) & 22.86 & 1.4 & \underline{23.53} & 13.77 & 9.2 & 17.79 & 16.28 & 7.4 & \underline{19.18} \\
 & USEF-TFGridNet (Static)   & 22.84 & 1.4 & 23.48 & 13.75 & 9.2 & 17.82 & 16.28 & 7.5 & 19.05 \\
 & EvoTSE ($k$=3)        & \underline{23.34} & \underline{0.6} & \underline{23.53} & \underline{16.23} & \underline{4.3} & \underline{17.85} & \underline{17.27} & \textbf{5.1} & 19.14 \\
 & EvoTSE ($k$=24)       & \textbf{23.39} & \textbf{0.5} & \textbf{23.56} & \textbf{16.67} & \textbf{3.5} & \textbf{17.90} & \textbf{17.43} & \textbf{5.1} & \textbf{19.23} \\ \bottomrule
\end{tabular}
\end{table*}

\subsection{Implementation Details}
\label{sec:ImplementationDetails}
\textbf{Model Configuration:} We adopt USEF-TFGridNet~\cite{USEFTSE} as the primary backbone. To ensure a fair comparison, we strictly follow the parameter settings from the original implementation. The model uses a window length of 16 ms and a hop length of 8 ms for STFT, extracting 65-dimensional complex features. The core architecture includes a 2D convolution layer with 128 output channels and a CMHA module with 4 parallel attention heads. Other hyperparameters, such as the 512-dimensional feed-forward network, are also kept consistent with the backbone’s reference configuration.

\textbf{Training Pipeline:} We apply a two-stage training strategy as follows: \textbf{Stage I }: The model is trained for up to 150 epochs with an initial learning rate of $5 \times 10^{-4}$, halved if the validation loss plateaus for 3 consecutive epochs. \textbf{Stage II }: We initialize the model with the Stage I checkpoint and perform Artifact-aware Curriculum Learning for 2 additional epochs using sequential chain-inference. For a fair comparison, the conventional baseline is also fine-tuned for an equivalent number of steps using standard static protocols.

\textbf{Inference Settings:} During evaluation, we compare our approach with two baseline inference modes: \textbf{Standard:} This mode follows the traditional protocol where each mixture is processed using a randomly selected enrollment utterance from the target speaker. \textbf{Static:} To align with our sequential setup, mixtures are grouped by speaker identity. A single initial enrollment is selected for each group, but the model extracts all segments using this fixed anchor without any updates. \textbf{EvoTSE:} Similar to the grouped baseline, we process mixtures in speaker-specific chains using a single initial enrollment. However, EvoTSE adaptively retrieves and integrates historical cues, as described in Section \ref{sec:method}.

\section{Experimental Results}
\subsection{Main Results}
Table \ref{tab:main_results} compares the performance of our proposed method with two baselines. USEF-TFGridNet (Standard) refers to the conventional inference pipeline. USEF-TFGridNet (Static) uses a grouped inference setup but keeps the enrollment fixed. EvoTSE represents our proposed grouped inference method with evolving updates. More details of these three modes can be found in Section \ref{sec:ImplementationDetails}. For EvoTSE, the parameter $k$ indicates the number of top segments retrieved from memory bank, which is analyzed further in Section \ref{sec:RetrievalQuantity}. This table summarizes the performance of models trained on two sets (WSJ or WSJ+ESD) and evaluated across three test sets (WSJ0-2mix, ESD-test, and Libri2mix-clean). For clarity, we adopt USEF-TFGridNet (Static) and EvoTSE ($k=3$) as the configuration for all remaining experiments, unless specified otherwise.

\textbf{Cross-Domain Generalization:} EvoTSE demonstrates superior OOD generalization compared to the baseline. Specifically, we evaluate two OOD scenarios: (a) training on WSJ and testing on ESD to assess performance under complex emotional variations, and (b) training on WSJ or WSJ+ESD and testing on Libri2mix-clean to evaluate generalization in standard acoustic environments. In scenario (a), USEF-TFGridNet (Static) achieves an SI-SDRi of 2.09 dB and an NSR of 23.9\%. In contrast, the EvoTSE model trained only on WSJ achieves an NSR of 8.1\%, which already outperforms the baseline trained on the WSJ+ESD dataset (9.2\%). Meanwhile, EvoTSE significantly improves the SI-SDRi from 2.09 dB to 10.73 dB. A similar trend is observed in scenario (b). For models trained on WSJ, EvoTSE improves the SI-SDRi from 16.65 dB to 17.91 dB. While EvoTSE remains superior when trained on WSJ+ESD (17.27 dB), both models show slightly lower performance than their WSJ-only counterparts. This is because the emotional variety in ESD, though beneficial for robustness, introduces additional acoustic variance that interferes with the model’s precision in the emotion-free environments.

\textbf{In-Domain Performance:} EvoTSE also shows strong results on in-domain test sets, especially in the more challenging emotional scenarios of ESD. We consider three in-domain scenarios: (a) training and testing on WSJ, (b) training on WSJ+ESD and testing on WSJ, and (c) training and testing on ESD. In scenario (a), EvoTSE maintains a low NSR of 0.5\%, outperforming the baseline’s 1.2\%. In scenario (b), the baseline's NSR worsens from 1.2\% to 1.4\%, as the diverse emotional data in WSJ+ESD may interfere with the model’s focus on the standard WSJ domain. However, EvoTSE relaxes this degradation and achieves a NSR of 0.6\%. Finally, in scenario (c), EvoTSE significantly improves the SI-SDRi from 13.75 dB to 16.23 dB and the NSR from 9.2\% to 4.3\%. 

\textbf{Consistency in SI-SDRiC:}
While EvoTSE significantly improves the overall SI-SDRi and NSR, the SI-SDRiC remains relatively stable across all methods. For example, when trained on WSJ, the SI-SDRiC for all methods on the ESD-test is approximately 13.5 dB, and when trained on WSJ+ESD, this value consistently reaches around 17.8 dB. These results indicate that the backbone extractor performs consistently once the target speaker is correctly identified. Therefore, EvoTSE does not specifically optimize the model's basic separation capability. Instead, it selects the most appropriate enrollments, which significantly mitigate  speaker confusion.

\subsection{Robustness to Initial Enrollment Variability}
To further investigate the sensitivity of EvoTSE to the quality of the initial enrollment, we evaluate the model's performance on ESD-test across five distinct emotional states of the initial enrollment, as shown in Table \ref{tab:emotional_robustness}.

\begin{table}[t]
\centering
\caption{Detailed performance on ESD-test categorized by the emotional state of the initial enrollment. USEF (Static) is short for USEF-TFGridNet (Static).}
\label{tab:emotional_robustness}

\setlength{\tabcolsep}{4pt} 
\resizebox{\columnwidth}{!}{%
\begin{tabular}{@{} l l ccc ccc @{}} 
\toprule
\textbf{Train} & \textbf{Init. Type} & \multicolumn{3}{c}{\textbf{USEF (Static)}} & \multicolumn{3}{c}{\textbf{EvoTSE (k=3)}} \\ 
\cmidrule(lr){3-5} \cmidrule(lr){6-8}
& & \makecell{\textbf{SI-SDRi}\\(dB)$\uparrow$} & \makecell{\textbf{NSR}\\(\%)$\downarrow$} & \makecell{\textbf{SI-SDRiC}\\(dB)$\uparrow$} & \makecell{\textbf{SI-SDRi}\\(dB)$\uparrow$} & \makecell{\textbf{NSR}\\(\%)$\downarrow$} & \makecell{\textbf{SI-SDRiC}\\(dB)$\uparrow$} \\ \midrule

\multirow{5}{*}{\makecell[l]{WSJ}} & Angry & -0.39 & 28.0 & 13.35 & \textbf{11.24} & \textbf{6.7} & \textbf{13.50} \\
 & Happy & -1.17 & 29.5 & 13.11 & \textbf{10.26} & \textbf{8.7} & \textbf{13.38} \\
 & Neutral & 5.90 & 16.6 & 13.32 & \textbf{10.43} & \textbf{8.9} & \textbf{13.48} \\
 & Sad & 6.21 & 16.6 & 13.48 & \textbf{10.86} & \textbf{8.2} & \textbf{13.56} \\
 & Surprise & -0.07 & 28.5 & \textbf{13.58} & \textbf{10.84} & \textbf{7.6} & 13.33 \\ \midrule

\multirow{5}{*}{\makecell[l]{WSJ + \\ ESD}} & Angry & 13.49 & 9.8 & 17.71 & \textbf{16.28} & \textbf{4.2} & \textbf{17.89} \\
 & Happy & 13.48 & 9.7 & 17.70 & \textbf{16.21} & \textbf{4.1} & \textbf{17.82} \\
 & Neutral & 15.05 & 6.7 & \textbf{17.98} & \textbf{16.09} & \textbf{4.6} & 17.85 \\
 & Sad & 14.95 & 6.9 & \textbf{17.95} & \textbf{16.42} & \textbf{4.0} & 17.91 \\
 & Surprise & 11.77 & 12.9 & 17.75 & \textbf{16.13} & \textbf{4.5} & \textbf{17.80} \\ \bottomrule
\end{tabular}%
}
\end{table}

\textbf{Acoustic Sensitivity of the Baseline:} The performance of the baseline model is heavily influenced by the quality of the initial enrollment. Using Neutral or Sad samples as the initial enrollment leads to much better results than using Angry, Happy or Surprise samples. This performance gap is especially large in the OOD scenario (Train on WSJ / Test on ESD), where the NSR for "Happy" (29.5\%) is nearly double that of "Neutral" (16.6\%). 

\textbf{Consistency and Robustness of EvoTSE:} EvoTSE exhibits remarkable stability regardless of the initial emotion, achieving significant improvements in both in-domain and OOD scenarios. Specifically, in the OOD case, EvoTSE consistently maintains an SI-SDRi of approximately 10.8 dB  regardless of the starting enrollment, effectively eliminating the severe performance drops seen in the baseline. For the in-domain scenario (Train on WSJ+ESD / Test on ESD), the performance remains stable at around 16.2 dB. This stability stems from the model's ability to refine the suboptimal enrollment through its evolution, which effectively cancels out the bias of the initial enrollment and leads to superior speaker identification.

\subsection{Ablation Study: Effectiveness of Auxiliary Information}
To further analyze the impact of enrollment, we evaluate several configurations on the ESD-test. As shown in Table \ref{tab:auxiliary_analysis}, "Static" represents the traditional baseline using a single fixed enrollment audio. "Multiple ($k$=24)" randomly selects 24 segments of the target speaker's speech and concatenating them in the time domain. We chose $k$=24 rather than $k$=3 because performance peaks at this point and slightly decreases if more segments are added. Finally, "EvoTSE ($k$=24)" shows our proposed method's performance when the retrieval size is set to 24. These configurations are also compared against the "Oracle Label", which uses the ground-truth target speech as the enrollment.

\textbf{Upper Bound Analysis using Oracle Labels:} The use of Oracle Labels as enrollment provides the best results across all metrics. Specifically, in both in-domain and OOD test sets, the NSR reaches a perfect or near-perfect state (0.0\% and 0.1\%). Along with these results, the SI-SDRiC also shows a noticeable improvement. In conclusion, the Oracle Label serves as the Top Line for methods that focus on optimizing auxiliary information, which is the ideal result we aim to reach by refining the enrollment process.

\textbf{Comparison Between Evolving Updates and Static Enrollments:} EvoTSE outperforms both the Static baseline across all metrics in both in-domain and OOD scenarios. When comparing Multi-Enroll to the Static baseline, the results show that simply increasing the number and variety of enrollments leads to better performance. For example, on the WSJ training set, Multi-Enroll reduces the NSR from 23.9\% to 17.0\%. However, our proposed EvoTSE achieves much better results than Multi-Enroll, especially on the OOD test set, where it further reduces the NSR down to 6.4\%. It is important to note that while Multi-Enroll requires multiple pre-existing clean reference audios, EvoTSE starts with only a single random enrollment. These findings demonstrate that our method is more effective at refining the enrollment.

\begin{table}[t]
\centering
\caption{Comparative analysis of auxiliary information types on ESD-test.}
\label{tab:auxiliary_analysis}

\setlength{\tabcolsep}{5pt} 

\resizebox{\columnwidth}{!}{%
\begin{tabular}{@{} l l ccc @{}} 
\toprule
\textbf{Train} & \textbf{Enrollment Type} & \textbf{SI-SDRi (dB) $\uparrow$} & \textbf{NSR (\%) $\downarrow$} & \textbf{SI-SDRiC (dB) $\uparrow$} \\ \midrule

\multirow{4}{*}{\makecell[l]{WSJ}} & USEF (Static)  & 2.09 & 23.9 & 13.37 \\
 & Multiple ($k$=24) & 5.59 & 17.0 & \underline{13.88} \\
 & EvoTSE ($k$=24) & \underline{11.34} & \underline{6.4} & 13.59 \\ \cmidrule(l){2-5}
 & Oracle Label & \textbf{14.95} & \textbf{0.1} & \textbf{14.97} \\ \midrule

\multirow{4}{*}{\makecell[l]{WSJ + \\ ESD}} & USEF (Static)  & 13.75 & 9.2 & 17.82 \\
 & Multiple ($k$=24) & 15.12 & 6.5 & \underline{17.95} \\
 & EvoTSE ($k$=24) & \underline{16.67} & \underline{3.5} & 17.90 \\ \cmidrule(l){2-5}
 & Oracle Label & \textbf{18.55} & \textbf{0.0} & \textbf{18.55} \\ \bottomrule
\end{tabular}%
}
\end{table}

\subsection{Ablation Study: Sensitivity of Similarity Threshold}
The Reliability Filter acts as a gatekeeper to ensure the purity of the memory bank (see Section \ref{sec:Classifier}). We investigate the impact of the similarity threshold $\tau$ on the ESD-test set to understand the trade-off between the quantity and quality of stored information, as shown in Fig. \ref{fig:threshold_uniform_plots}. The horizontal dashed lines represent the baseline results for models trained on WSJ+ESD (red) and WSJ-only (blue). 

\begin{figure*}[t!]
    \centering
    \newcommand{\tauwidth}{6.0cm}  
    \newcommand{\tauheight}{5.5cm} 

    \pgfplotsset{
        threshold_uniform_plot/.style={
            width=\tauwidth,
            height=\tauheight,
            symbolic x coords={0.0, 0.2, 0.4, 0.5, 0.6, 0.8, 1.0},
            xtick=data,
            grid=major,
            grid style={dashed, gray!30},
            nodes near coords,
            every node near coord/.append style={font=\tiny, anchor=south},
            label style={font=\small},
            tick label style={font=\small}
        }
    }

    \begin{subfigure}{0.32\textwidth}
        \centering
        \begin{tikzpicture}
            \begin{axis}[threshold_uniform_plot, ylabel={SISDRi (dB)}, ymin=0, ymax=18.5]
                \draw[blue!80, dashed, thick] (axis cs:0.0, 2.09) -- (axis cs:1.0, 2.09);
                \draw[red!80, dashed, thick] (axis cs:0.0, 13.75) -- (axis cs:1.0, 13.75);
                
                \addplot[blue!70, mark=square*, thick] coordinates {
                    (0.0, 1.03) (0.2, 4.80) (0.4, 8.90) (0.5, 10.73) (0.6, 9.99) (0.8, 5.50) (1.0, 2.09)
                };
                \addplot[red!80, mark=triangle*, thick] coordinates {
                    (0.0, 11.30) (0.2, 13.10) (0.4, 15.20) (0.5, 16.23) (0.6, 16.12) (0.8, 14.80) (1.0, 13.75)
                };
            \end{axis}
        \end{tikzpicture}
        \caption{SISDRi across $\tau$}
    \end{subfigure}
    \hfill
    \begin{subfigure}{0.32\textwidth}
        \centering
        \begin{tikzpicture}
            \begin{axis}[threshold_uniform_plot, ylabel={NSR (\%)}, ymin=0, ymax=30]
                \draw[blue!70, dashed, thick] (axis cs:0.0, 23.9) -- (axis cs:1.0, 23.9);
                \draw[red!80, dashed, thick] (axis cs:0.0, 9.2) -- (axis cs:1.0, 9.2);
    
                \addplot[blue!70, mark=square*, thick] coordinates {
                    (0.0, 26.8) (0.2, 19.2) (0.4, 12.5) (0.5, 8.1) (0.6, 10.1) (0.8, 16.5) (1.0, 23.9)
                };
                \addplot[red!80, mark=triangle*, thick] coordinates {
                    (0.0, 14.3) (0.2, 10.8) (0.4, 6.5) (0.5, 4.3) (0.6, 4.7) (0.8, 7.2) (1.0, 9.2)
                };
            \end{axis}
        \end{tikzpicture}
        \caption{NSR across $\tau$}
    \end{subfigure}
    \hfill
    \begin{subfigure}{0.32\textwidth}
        \centering
        \begin{tikzpicture}
            \begin{axis}[threshold_uniform_plot, ylabel={SISDRiC (dB)}, ymin=12, ymax=19.5]

                \draw[blue!70, dashed, thick] (axis cs:0.0, 17.82) -- (axis cs:1.0, 17.82);
                \draw[red!80, dashed, thick] (axis cs:0.0, 13.37) -- (axis cs:1.0, 13.37);
                \draw[blue!70, dashed, thick] (axis cs:0.0, 10) -- (axis cs:1.0, 10)
                    node[pos=0, anchor=south west, font=\tiny, xshift=2pt] {10.0};
                \draw[red!80, dashed, thick] (axis cs:0.0, 10) -- (axis cs:1.0, 10)
                    node[pos=0, anchor=south west, font=\tiny, xshift=2pt] {10.0};

                \addplot[blue!70, mark=square*, thick] coordinates {
                    (0.0, 13.43) (0.2, 13.40) (0.4, 13.42) (0.5, 13.45) (0.6, 13.68) (0.8, 13.50) (1.0, 13.37)
                };
                \addplot[red!80, mark=triangle*, thick] coordinates {
                    (0.0, 17.17) (0.2, 17.50) (0.4, 17.70) (0.5, 17.85) (0.6, 17.82) (0.8, 17.80) (1.0, 17.82)
                };
            \end{axis}
        \end{tikzpicture}
        \caption{SISDRiC across $\tau$}
    \end{subfigure}

    \caption{Performance metrics as a function of similarity threshold $\tau$ on ESD-test ($k=3$, $|\mathcal{M}|_{\max}=64$). Dashed lines represent USEF-TFGridnet results. Red triangle lines and blue square lines distinguish models trained on WSJ+ESD and WSJ, respectively.}
    \label{fig:threshold_uniform_plots}
\end{figure*}
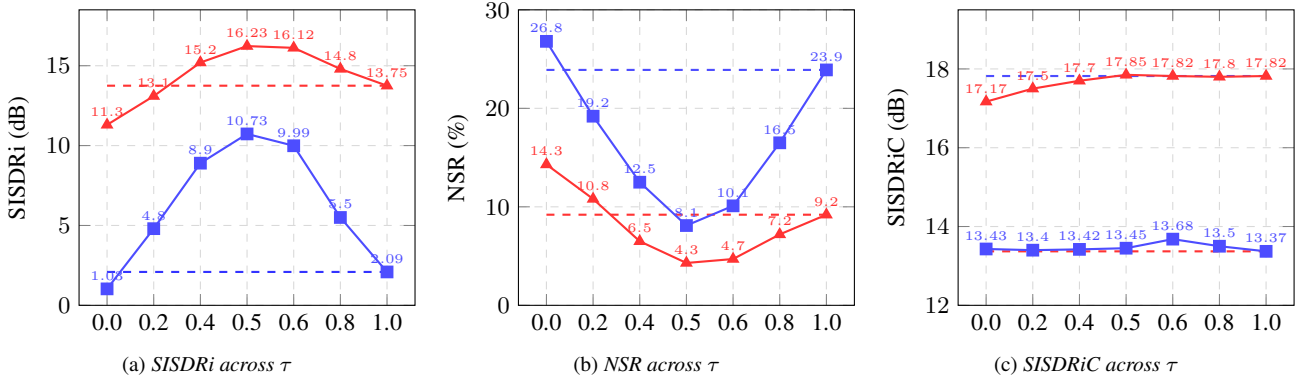

\begin{figure*}[t!]
    \centering
    \newcommand{\figwidth}{5.8cm}  
    \newcommand{\figheight}{5.5cm} 

    \pgfplotsset{
        my_line_plot/.style={
            width=\figwidth,
            height=\figheight,
            symbolic x coords={1, 3, 12, 24, 48, 64}, 
            xtick=data,
            grid=major,
            grid style={dashed, gray!30},
            nodes near coords,
            every node near coord/.append style={font=\tiny, anchor=south},
            label style={font=\small},
            tick label style={font=\small}
        }
    }

    \begin{subfigure}{0.32\textwidth}
        \centering
        \begin{tikzpicture}
            \begin{axis}[my_line_plot, ylabel={SISDRi(dB)}, ymin=8, ymax=18.5]
                \addplot[blue!70, mark=square*, thick] coordinates {(1,10.09) (3,10.73) (12,11.63) (24,11.34) (48,9.78) (64,9.09)};
                \addplot[red!80, mark=triangle*, thick] coordinates {(1,15.14) (3,16.23) (12,16.63) (24,16.67) (48,16.36) (64,16.32)};
            \end{axis}
        \end{tikzpicture}
        \caption{SISDRi across k}
    \end{subfigure}
    \hfill
    \begin{subfigure}{0.32\textwidth}
        \centering
        \begin{tikzpicture}
            \begin{axis}[my_line_plot, ylabel={NSR (\%)}, ymin=0, ymax=15]
                \addplot[blue!70, mark=square*, thick] coordinates {(1,9.6) (3,8.1) (12,5.9) (24,6.4) (48,10.3) (64,11.6)};
                \addplot[red!80, mark=triangle*, thick] coordinates {(1,6.8) (3,4.3) (12,3.7) (24,3.5) (48,4.1) (64,4.1)};
            \end{axis}
        \end{tikzpicture}
        \caption{NSR across k}
    \end{subfigure}
    \hfill
    \begin{subfigure}{0.32\textwidth}
        \centering
        \begin{tikzpicture}
            \begin{axis}[my_line_plot, ylabel={SISDRiC(dB)}, ymin=13, ymax=19]
                \addplot[blue!70, mark=square*, thick] coordinates {(1,13.46) (3,13.45) (12,13.66) (24,13.59) (48,13.42) (64,13.40)};
                \addplot[red!80, mark=triangle*, thick] coordinates {(1,17.83) (3,17.85) (12,17.98) (24,17.90) (48,17.86) (64,17.83)};
            \end{axis}
        \end{tikzpicture}
        \caption{SISDRiC across k}
    \end{subfigure}

    \caption{Performance metrics as a function of retrieval quantity $k$ on ESD-test  ($\tau=0.5$, $|\mathcal{M}|_{\max}=64$). Red triangle lines indicate models trained on WSJ + ESD, while blue square lines represent models trained on WSJ only.}
    \label{fig:metrics_all}
\end{figure*}
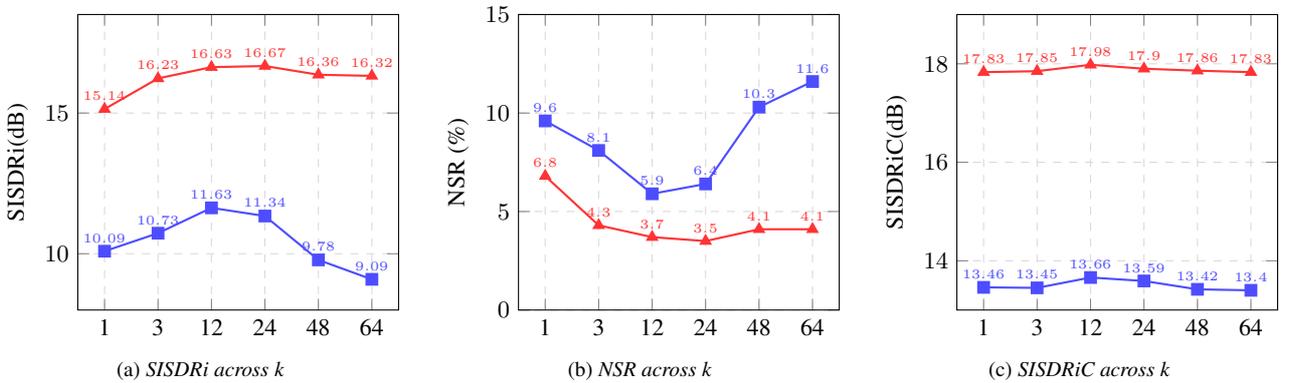

\textbf{Memory Poisoning at $\tau = 0.0$:} When the threshold is set to $\tau = 0.0$, the Reliability Filter is disabled, and every extraction result is stored in the memory bank regardless of its quality. Quantitatively, this lack of filtering leads to a sharp performance decline that falls even below the baseline in some cases. Trained on the WSJ+ESD set, the NSR increases to 14.3\%, which is much higher than the 9.2\% baseline. Similarly, trained on the WSJ set, the SI-SDRi drops to 1.03 dB, failing to reach the 2.09 dB baseline. These results demonstrate that without constraints, EvoTSE suffers from memory poisoning, where incorrect segments from interfering speakers are captured and amplified.

\textbf{Optimal Balance at $\tau = 0.5$:} As the threshold $\tau$ increases, the requirements for adding new segments to the memory bank become more strict. When $\tau = 1.0$, EvoTSE performs exactly like the staic baseline. The best results are achieved at $\tau = 0.5$, which balances the need for new information with the need for high quality. For the model trained on WSJ, the NSR improves from 23.9\% to 8.1\%, while the SI-SDRi increases from 2.09 dB to 10.73 dB. This gain is more noticeable in OOD scenarios, showing that the threshold is critical for correcting identity drift in unfamiliar environments. In conclusion, the consistent success across settings proves that the EvoTSE effectively filtering out interfering while capturing necessary variations.

\subsection{Ablation Study: Impact of Retrieval Quantity}
\label{sec:RetrievalQuantity}
We examine the influence of the retrieval quantity $k$ to understand how the amount of historical evidence affects the extraction process. As shown in Fig. \ref{fig:metrics_all}, the results reveals that more information does not always lead to better results.


\textbf{Superiority of Top-$k$ Retrieval:} The results demonstrate that strong performance can be achieved with only a small number of retrieved segments. As shown in Fig. \ref{fig:metrics_all}, even at $k=1$, EvoTSE already outperforms the baseline. For instance, the SI-SDRi for the WSJ-only model starts at 10.09 dB, which is significantly higher than its 2.09 dB baseline. As $k$ increases, performance continues to improve and peaks between $k=12$ and $24$, where the NSR for the WSJ+ESD model reaches its lowest point of 3.5\%. This shows that selecting only a few highly relevant enrollments from the memory bank is enough to bring a significant improvement to the extraction quality.

\textbf{Degradation of Global Aggregation:} When $k$ increases further, providing more historical information does not necessarily lead to better performance. As shown in Fig. \ref{fig:metrics_all}, when $k$ reaches 64 (utilizing the entire memory bank), both in-domain and OOD metrics exhibit a notable decline. This degradation is especially clear in the OOD (WSJ-only) scenario, where the SI-SDRi drops from 11.63 dB to 9.09 dB. These results confirm that our strategy of selecting only the Locally Relevant Context (Top-$k$) is more effective than using the entire history.

\begin{table}[t]
\centering
\caption{Impact of alignment fine-tuning on the ESD-test. "FT" denotes the Stage-II fine-tuning.}
\label{tab:ft_ablation}

\setlength{\tabcolsep}{5pt} 

\resizebox{\columnwidth}{!}{
\begin{tabular}{@{} l l ccc @{}} 
\toprule
\textbf{Train} & \textbf{Method} & \textbf{SI-SDRi (dB) $\uparrow$} & \textbf{NSR (\%) $\downarrow$} & \textbf{SI-SDRiC (dB) $\uparrow$} \\ \midrule

\multirow{3}{*}{\makecell[l]{WSJ}} & USEF (Static)  & 2.09 & 23.9 & 13.37 \\
 & EvoTSE (w/o FT) & 1.55 & 30.2 & 13.10 \\
 & EvoTSE (w/ FT) & \textbf{10.73} & \textbf{8.1} & \textbf{13.45} \\ \midrule

\multirow{3}{*}{\makecell[l]{WSJ + \\ ESD}} & USEF (Static)  & 13.75 & 9.2 & 17.82 \\
 & EvoTSE (w/o FT) & 10.40 & 15.8 & 17.61 \\
 & EvoTSE (w/ FT) & \textbf{16.23} & \textbf{4.3} & \textbf{17.85} \\ \bottomrule
\end{tabular}%
}
\end{table}

\subsection{Ablation Study: Efficacy of Artifact-aware Learning}
As shown in Table \ref{tab:ft_ablation}, to verify the necessity of the proposed two-stage training strategy, we compare the baseline with two variants: one without Stage-II fine-tuning (Wo FT) and one with the full fine-tuning (Wi FT).

\textbf{Artifact Mitigation via Stage-II Learning}: Stage-II training is essential for bridging the artifact gap between ideal training conditions and real-world evolving updates. When the Stage-I model is used directly without fine-tuning (w/o FT), the NSR degrades significantly from 9.2\% to 15.8\%. This performance drop occurs because the initial model was trained only on clean enrollment signals. In the evolving loop, the updated enrollment inevitably contains subtle neural processing artifacts and residual interfering, leading to identity confusion. However, by introducing the Artifact-aware Learning (w/ FT), EvoTSE learns to extract stable identity cues even from imperfect enrollments. This specialized training effectively closes the loop and further reduces the NSR to 4.3\%. These results prove that fine-tuning is necessary to ensure the model can handle the signal-level imperfections inherent in a self-updating system.

\section{Conclusions}
This paper presents EvoTSE, a framework that transitions TSE from static mapping to an evolving inference pipeline by evolving the enrollment representation. Our approach effectively mitigates speaker confusion, particularly in complex OOD scenarios, and significantly reduces dependency on the quality of the initial enrollment audio. However, the dynamic retrieval and memory evolution mechanisms introduce additional computational overhead and inference latency. Future work will focus on optimizing these components to balance performance gains with real-time deployment requirements.

\section{Generative AI Use Disclosure}
During the preparation of this manuscript, the authors utilized large language models (LLMs) solely for grammatical verification and language refinement. Specifically, these tools were employed to ensure grammatical consistency, improve sentence fluency, and enhance overall readability. The authors emphasize that LLMs played no role in the core scientific components of this study, including algorithm design, data processing, model training, or experimental evaluation. The final content was thoroughly reviewed and edited by the authors, who take full responsibility for the integrity and accuracy of the work.
\bibliographystyle{IEEEtran}
\bibliography{mybib}

@INPROCEEDINGS{CSS,
  author={Chen, Zhuo and Yoshioka, Takuya and Lu, Liang and Zhou, Tianyan and Meng, Zhong and Luo, Yi and Wu, Jian and Xiao, Xiong and Li, Jinyu},
  booktitle={IEEE International Conference on Acoustics, Speech and Signal Processing (ICASSP)}, 
  title={Continuous Speech Separation: Dataset and Analysis}, 
  year={2020},
  volume={},
  number={},
  pages={7284-7288},
  doi={10.1109/ICASSP40776.2020.9053426}}

@ARTICLE{USEFTSE,
  author={Zeng, Bang and Li, Ming},
  journal={IEEE Transactions on Audio, Speech and Language Processing}, 
  title={USEF-TSE: Universal Speaker Embedding Free Target Speaker Extraction}, 
  year={2025},
  volume={33},
  number={},
  pages={2110-2124},
  doi={10.1109/TASLPRO.2025.3572756}
}

@inproceedings{bsrnn_tfmap,
author = {Zhang, Ke and Li, Junjie and Wang, Shuai and Wei, Yangjie and Wang, Yi and Wang, Yannan and Li, Haizhou},
year = {2025},
month = {04},
pages = {1-5},
booktitle={IEEE International Conference on Acoustics, Speech and Signal Processing (ICASSP)},
title = {Multi-Level Speaker Representation for Target Speaker Extraction},
doi = {10.1109/ICASSP49660.2025.10889409}
}

@article{spex,
author = {Xu, Chenglin and Rao, Wei and Chng, Eng and Li, Haizhou},
year = {2020},
month = {04},
pages = {1-1},
title = {SpEx: Multi-Scale Time Domain Speaker Extraction Network},
volume = {PP},
journal = {IEEE/ACM Transactions on Audio, Speech, and Language Processing},
doi = {10.1109/TASLP.2020.2987429}
}

@inproceedings{spexplus,
author = {Ge, Meng and Xu, Chenglin and Wang, Longbiao and Chng, Eng and Dang, Jianwu and Li, Haizhou},
year = {2020},
month = {10},
pages = {1406-1410},
title = {SpEx+: A Complete Time Domain Speaker Extraction Network},
booktitle = {Proc. INTERSPEECH 2020},
doi = {10.21437/Interspeech.2020-1397}
}

@inproceedings{voicefilter,
author = {Wang, Quan and Muckenhirn, Hannah and Wilson, Kevin and Sridhar, Prashant and Wu, Zelin and Hershey, John and Saurous, Rif and Weiss, Ron and Jia, Ye and Moreno, Ignacio},
year = {2019},
month = {09},
pages = {2728-2732},
title = {VoiceFilter: Targeted Voice Separation by Speaker-Conditioned Spectrogram Masking},
booktitle    = {Proc. INTERSPEECH 2019},
doi = {10.21437/Interspeech.2019-1101}
}

@inproceedings{Wang2018DeepEN,
author = {Wang, Jun and Chen, Jie and Su, Dan and Chen, Lianwu and Yu, Meng and Qian, Yanmin and Yu, Dong},
year = {2018},
month = {09},
pages = {307-311},
title = {Deep Extractor Network for Target Speaker Recovery from Single Channel Speech Mixtures},
booktitle={Proc. INTERSPEECH 2018},
doi = {10.21437/Interspeech.2018-1205}
}

@article{speakerbeam,
author = {Žmolíková, Kateřina and Delcroix, Marc and Kinoshita, Keisuke and Ochiai, Tsubasa and Nakatani, Tomohiro and Burget, Lukas and Cernocky, Jan},
year = {2019},
month = {06},
pages = {1-1},
title = {SpeakerBeam: Speaker Aware Neural Network for Target Speaker Extraction in Speech Mixtures},
volume = {PP},
journal = {IEEE Journal of Selected Topics in Signal Processing},
doi = {10.1109/JSTSP.2019.2922820}
}

@INPROCEEDINGS{TEAPSE,
  author={Ju, Yukai and Rao, Wei and Yan, Xiaopeng and Fu, Yihui and Lv, Shubo and Cheng, Luyao and Wang, Yannan and Xie, Lei and Shang, Shidong},
  
  booktitle={IEEE International Conference on Acoustics, Speech and Signal Processing (ICASSP)},
  title={TEA-PSE: Tencent-Ethereal-Audio-Lab Personalized Speech Enhancement System for ICASSP 2022 DNS Challenge}, 
  year={2022},
  volume={},
  number={},
  pages={9291-9295},
  doi={10.1109/ICASSP43922.2022.9747765}}

@INPROCEEDINGS{TEAPSE2,
  author={Ju, Yukai and Zhang, Shimin and Rao, Wei and Wang, Yannan and Yu, Tao and Xie, Lei and Shang, Shidong},
  booktitle={2022 IEEE Spoken Language Technology Workshop (SLT)}, 
  title={TEA-PSE 2.0: Sub-Band Network for Real-Time Personalized Speech Enhancement}, 
  year={2023},
  volume={},
  number={},
  pages={472-479},
  doi={10.1109/SLT54892.2023.10023174}}

@INPROCEEDINGS{TEAPSE3,
  author={Ju, Yukai and Chen, Jun and Zhang, Shimin and He, Shulin and Rao, Wei and Zhu, Weixin and Wang, Yannan and Yu, Tao and Shang, Shidong},
  booktitle={IEEE International Conference on Acoustics, Speech and Signal Processing (ICASSP)},
  title={TEA-PSE 3.0: Tencent-Ethereal-Audio-Lab Personalized Speech Enhancement System For ICASSP 2023 Dns-Challenge}, 
  year={2023},
  volume={},
  number={},
  pages={1-2},
  doi={10.1109/ICASSP49357.2023.10096838}}

@article{HAO2024102550,
title = {X-TF-GridNet: A time–frequency domain target speaker extraction network with adaptive speaker embedding fusion},
journal = {Information Fusion},
volume = {112},
pages = {102550},
year = {2024},
issn = {1566-2535},
doi = {https://doi.org/10.1016/j.inffus.2024.102550},
author = {Fengyuan Hao and Xiaodong Li and Chengshi Zheng},

}

@article{Shulin,
author = {He, Shulin and Xue, Wei and Yang, Yang and Zhang, Huaiwen and Pan, Jiahao and Zhang, Xueliang},
title = {Enhancing target speaker extraction with Hierarchical Speaker Representation Learning},
year = {2025},
volume = {188},
issn = {0893-6080},
url = {https://doi.org/10.1016/j.neunet.2025.107388},
doi = {10.1016/j.neunet.2025.107388},
journal = {Neural Networks},
month = {8},
numpages = {11},
}

@INPROCEEDINGS{HeShulin,
  author={He, Shulin and Zhang, Huaiwen and Rao, Wei and Zhang, Kanghao and Ju, Yukai and Yang, Yang and Zhang, Xueliang},
  booktitle={IEEE International Conference on Acoustics, Speech and Signal Processing (ICASSP)}, 
  title={Hierarchical Speaker Representation for Target Speaker Extraction}, 
  year={2024},
  volume={},
  number={},
  pages={10361-10365},
  doi={10.1109/ICASSP48485.2024.10447755}}

@INPROCEEDINGS{HeShulin20,
  author={He, Shulin and Li, Hao and Zhang, Xueliang},
  booktitle={IEEE International Conference on Acoustics, Speech and Signal Processing (ICASSP)}, 
  title={Speakerfilter: Deep Learning-Based Target Speaker Extraction Using Anchor Speech}, 
  year={2020},
  volume={},
  number={},
  pages={376-380},
  doi={10.1109/ICASSP40776.2020.9054222}}

@inproceedings{TargetConfusion,
author = {Zhao, Zifeng and Yang, Dongchao and Rongzhi, Gu and Zhang, Haoran and Zou, Yuexian},
year = {2022},
month = {09.},
pages = {5333-5337},
title = {Target Confusion in End-to-end Speaker Extraction: Analysis and Approaches},
booktitle  = {Proc. INTERSPEECH 2022},
doi = {10.21437/Interspeech.2022-176}
}

@inproceedings{xsep_chunkloss,
author = {Liu, Kai and Du, Ziqing and Wan, Xucheng and Zhou, Huan},
year = {2023},
month = {06},
pages = {1-5},
title = {X-SEPFORMER: End-To-End Speaker Extraction Network with Explicit Optimization on Speaker Confusion},
booktitle={IEEE International Conference on Acoustics, Speech and Signal Processing (ICASSP)},
doi = {10.1109/ICASSP49357.2023.10095609}
}

@inproceedings{TargetConfusionAug,
author = {You, Zhenghai and Zhou, Zhenyu and Li, Lantian and Wang, Dong},
title     = {SpkAugTSE: A Simple and Efficient Approach to Address Target Confusion in End-to-End Speaker Extraction},
booktitle = {Proc. APSIPA ASC},
year      = {2025},
pages     = {583--588},
doi       = {10.1109/APSIPAASC65261.2025.11249310}
}

@inproceedings{tse_sv,
author = {Delcroix, Marc and Kinoshita, Keisuke and Ochiai, Tsubasa and Žmolíková, Kateřina and Sato, Hiroshi and Nakatani, Tomohiro},
year = {2022},
month = {09},
pages = {216-220},
title = {Listen only to me! How well can target speech extraction handle false alarms?},
booktitle    = {Proc. INTERSPEECH 2022},
doi = {10.21437/Interspeech.2022-11252}
}

@inproceedings{refine,
author = {Wang, Jiahe and Wang, Shuai and Li, Junjie and Zhang, Ke and Qian, Yanmin and Li, Haizhou},
year = {2024},
month = {12},
pages = {349-356},
title = {Enhancing Speaker Extraction Through Rectifying Target Confusion},
booktitle={2024 IEEE Spoken Language Technology Workshop (SLT)}, 
doi = {10.1109/SLT61566.2024.10832179}
}

@inproceedings{EnrollAug,
author = {Li, Junjie and Zhang, Ke and Wang, Shuai and Li, Haizhou and Mak, Man-Wai and Lee, Kong Aik},
year = {2024},
month = {12},
pages = {325-332},
title = {On the Effectiveness of Enrollment Speech Augmentation For Target Speaker Extraction},
booktitle={2024 IEEE Spoken Language Technology Workshop (SLT)}, 
doi = {10.1109/SLT61566.2024.10832217}
}

@INPROCEEDINGS{shortenroll,
  author={Yang, Lei and Liu, Wei and Tan, Lufen and Yang, Jaemo and Moon, Han-Gil},
  booktitle={IEEE International Conference on Acoustics, Speech and Signal Processing (ICASSP)}, 
  title={Target Speaker Extraction with Ultra-Short Reference Speech by VE-VE Framework}, 
  year={2023},
  volume={},
  number={},
  pages={1-5},
  doi={10.1109/ICASSP49357.2023.10096664}}

@inproceedings{DPRNNIRA,
author = {Chengyun, Deng and Ma, Shiqian and Sha, Yongtao and Zhang, Yi and Zhang, Hui and Song, Hui and Wang, Fei},
year = {2021},
month = {08},
pages = {3530-3534},
booktitle = {Proc. INTERSPEECH 2021},
title = {Robust Speaker Extraction Network Based on Iterative Refined Adaptation},
doi = {10.21437/Interspeech.2021-2250}
}

@inproceedings{SpEx++,
author = {Ge, Meng and Xu, Chenglin and Wang, Longbiao and Chng, Eng and Dang, Jianwu and Li, Haizhou},
year = {2021},
month = {06},
pages = {6109-6113},
booktitle={IEEE International Conference on Acoustics, Speech and Signal Processing (ICASSP)},
title = {Multi-Stage Speaker Extraction with Utterance and Frame-Level Reference Signals},
doi = {10.1109/ICASSP39728.2021.9413359}
}

@inproceedings{ortse,
author = {Zhang, Yiru and Yao, Linyu and Yang, Qun},
year = {2024},
month = {09},
pages = {587-591},
title = {OR-TSE: An Overlap-Robust Speaker Encoder for Target Speech Extraction},
booktitle = {Proc. INTERSPEECH 2024},
doi = {10.21437/Interspeech.2024-2322}
}

@misc{compareenroll,
      title={Target Speaker Extraction through Comparing Noisy Positive and Negative Audio Enrollments}, 
      author={Shitong Xu and Yiyuan Yang and Niki Trigoni and Andrew Markham},
      year={2025},
      eprint={2502.16611},
      archivePrefix={arXiv},
      primaryClass={cs.SD},
      url={https://arxiv.org/abs/2502.16611}, 
}

@ARTICLE{GhaneEnroll,
  author={Ghane, Mohsen and Safari, Mohammad Sadegh},
  journal={IEEE Signal Processing Letters}, 
  title={End-to-End Target Speaker Speech Recognition Using Context-Aware Attention Mechanisms for Challenging Enrollment Scenario}, 
  year={2025},
  volume={32},
  number={},
  pages={1940-1944},
}

@INPROCEEDINGS{SADenroll,
  author={Delcroix, Marc and Zmolikova, Katerina and Ochiai, Tsubasa and Kinoshita, Keisuke and Nakatani, Tomohiro},
  booktitle={IEEE International Conference on Acoustics, Speech and Signal Processing (ICASSP)},
  title={Speaker Activity Driven Neural Speech Extraction}, 
  year={2021},
  volume={},
  number={},
  pages={6099-6103},
  doi={10.1109/ICASSP39728.2021.9414998}}

@InProceedings{Nonoverlapp,
author="Zhang, Yiru
and Li, Zeke
and Liu, Bijing
and Fan, Haiwei
and Yang, Yong
and Yang, Qun",
title="A Region Based Non-overlapping Reference Speech Estimation Method for Speaker Extraction",
booktitle="MultiMedia Modeling",
year="2024",
publisher="Springer Nature Switzerland",
pages="437--447",
isbn="978-3-031-53311-2"
}

@article{Veluri2024LookOT,
  title={Look Once to Hear: Target Speech Hearing with Noisy Examples},
  author={Bandhav Veluri and Malek Itani and Tuochao Chen and Takuya Yoshioka and Shyamnath Gollakota},
  journal={Proceedings of the 2024 CHI Conference on Human Factors in Computing Systems},
  year={2024},
}

@article{TSsep,
author = {Boeddeker, Christoph and Subramanian, Aswin and Wichern, Gordon and Haeb-Umbach, Reinhold and Le Roux, Jonathan},
year = {2024},
month = {01},
pages = {1-13},
title = {TS-SEP: Joint Diarization and Separation Conditioned on Estimated Speaker Embeddings},
volume = {PP},
journal = {IEEE/ACM Transactions on Audio, Speech, and Language Processing},
doi = {10.1109/TASLP.2024.3350887}
}

@inproceedings{RAG,
author = {Lewis, Patrick and Perez, Ethan and Piktus, Aleksandra and Petroni, Fabio and Karpukhin, Vladimir and Goyal, Naman and K\"{u}ttler, Heinrich and Lewis, Mike and Yih, Wen-tau and Rockt\"{a}schel, Tim and Riedel, Sebastian and Kiela, Douwe},
title = {Retrieval-augmented generation for knowledge-intensive NLP tasks},
year = {2020},
isbn = {9781713829546},
address = {Red Hook, NY, USA},
booktitle = {Advances in Neural Information Processing Systems},
articleno = {793},
numpages = {16},
location = {Vancouver, BC, Canada},
}

@inproceedings{WavRAG,
    title = "WavRAG: Audio-Integrated Retrieval Augmented Generation for Spoken Dialogue Models",
    author = "Chen, Yifu  and
      Ji, Shengpeng  and
      Wang, Haoxiao  and
      Wang, Ziqing  and
      Chen, Siyu  and
      He, Jinzheng  and
      Xu, Jin  and
      Zhao, Zhou",
    booktitle = "Proceedings of the 63rd Annual Meeting of the Association for Computational Linguistics (Volume 1: Long Papers)",
    month = jul,
    year = "2025",
    address = "Vienna, Austria",
    publisher = "Association for Computational Linguistics",
    doi = "10.18653/v1/2025.acl-long.613",
    pages = "12505--12523",
    ISBN = "979-8-89176-251-0"
}

@INPROCEEDINGS{TTARag1,
  author={Yuan, Yi and Liu, Haohe and Liu, Xubo and Huang, Qiushi and Plumbley, Mark D. and Wang, Wenwu},
  
  booktitle={IEEE International Conference on Acoustics, Speech and Signal Processing (ICASSP)},
  title={Retrieval-Augmented Text-to-Audio Generation}, 
  year={2024},
  volume={},
  number={},
  pages={581-585},
  doi={10.1109/ICASSP48485.2024.10447898}}

@inproceedings{CAPRAG,
author = {Ghosh, Sreyan and Kumar, Sonal and Evuru, Chandra and Duraiswami, Ramani and Manocha, Dinesh},
year = {2024},
month = {04},
pages = {1161-1165},
title = {Recap: Retrieval-Augmented Audio Captioning},
booktitle={IEEE International Conference on Acoustics, Speech and Signal Processing (ICASSP)},
doi = {10.1109/ICASSP48485.2024.10448030}
}

@misc{garofolo1993csr,
  title        = {CSR-I (WSJ0) Complete LDC93S6A},
  author       = {Garofolo, John and Graff, David and Paul, Douglas and Pallett, David},
  year         = {1993},
  howpublished = {Web Download},
  publisher    = {Linguistic Data Consortium},
  address      = {Philadelphia},
  note         = {LDC93S6A}
}

@inproceedings{zhou2021seen,
  title={Seen and unseen emotional style transfer for voice conversion with a new emotional speech dataset},
  author={Zhou, Kun and Sisman, Berrak and Liu, Rui and Li, Haizhou},
  booktitle={IEEE International Conference on Acoustics, Speech and Signal Processing (ICASSP)},
  pages={920--924},
  year={2021},
  organization={IEEE}
}

@article{zhou2022emotional,
  title={Emotional voice conversion: Theory, databases and esd},
  author={Zhou, Kun and Sisman, Berrak and Liu, Rui and Li, Haizhou},
  journal={Speech Communication},
  volume={137},
  pages={1--18},
  year={2022},
  publisher={Elsevier}
}

@article{Cosentino2020LibriMixAO,
  title={LibriMix: An Open-Source Dataset for Generalizable Speech Separation},
  author={Joris Cosentino and Manuel Pariente and Samuele Cornell and Antoine Deleforge and Emmanuel Vincent},
  journal={arXiv: Audio and Speech Processing},
  year={2020},
  url={https://api.semanticscholar.org/CorpusID:218862876}
}

@INPROCEEDINGS{dc,
  author={Hershey, John R. and Chen, Zhuo and Le Roux, Jonathan and Watanabe, Shinji},
  booktitle={IEEE International Conference on Acoustics, Speech and Signal Processing (ICASSP)},
  title={Deep clustering: Discriminative embeddings for segmentation and separation}, 
  year={2016},
  volume={},
  number={},
  pages={31-35},
  doi={10.1109/ICASSP.2016.7471631}}

@inproceedings{xtas,
author = {Zhang, Zining and He, Bingsheng and Zhang, Zhenjie},
year = {2020},
month = {10},
pages = {1421-1425},
title = {X-TaSNet: Robust and Accurate Time-Domain Speaker Extraction Network},
booktitle = {Proc. INTERSPEECH 2020},
doi = {10.21437/Interspeech.2020-1706}
}

@article{tfgrid,
author = {Wang, Zhong-Qiu and Cornell, Samuele and Choi, Shukjae and Lee, Younglo and Kim, Byeong-Yeol and Watanabe, Shinji},
year = {2023},
month = {01},
pages = {1-15},
title = {TF-GridNet: Integrating Full- and Sub-Band Modeling for Speech Separation},
volume = {PP},
journal = {IEEE/ACM Transactions on Audio, Speech, and Language Processing},
doi = {10.1109/TASLP.2023.3304482}
}

@inproceedings{ecapa,
author = {Desplanques, Brecht and Thienpondt, Jenthe and Demuynck, Kris},
year = {2020},
month = {10},
pages = {},
title = {ECAPA-TDNN: Emphasized Channel Attention, Propagation and Aggregation in TDNN Based Speaker Verification},
booktitle    = {Proc. INTERSPEECH 2020},
doi = {10.21437/Interspeech.2020-2650}
}

@inproceedings{wang2023wespeaker,
  title={Wespeaker: A research and production oriented speaker embedding learning toolkit},
  author={Wang, Hongji and Liang, Chengdong and Wang, Shuai and Chen, Zhengyang and Zhang, Binbin and Xiang, Xu and Deng, Yanlei and Qian, Yanmin},
  booktitle={IEEE International Conference on Acoustics, Speech and Signal Processing (ICASSP)},
  pages={1--5},
  year={2023},
  organization={IEEE}
}

@inproceedings{ma2023emotion2vec,
    title = "emotion2vec: Self-Supervised Pre-Training for Speech Emotion Representation",
    author = "Ma, Ziyang  and
      Zheng, Zhisheng  and
      Ye, Jiaxin  and
      Li, Jinchao  and
      Gao, Zhifu  and
      Zhang, ShiLiang  and
      Chen, Xie",
    booktitle = "Findings of the Association for Computational Linguistics: ACL 2024",
    year = "2024",
    address = "Bangkok, Thailand",
    pages = "15747--15760",
}

@INPROCEEDINGS{muse,
  author={Pan, Zexu and Tao, Ruijie and Xu, Chenglin and Li, Haizhou},
  booktitle={IEEE International Conference on Acoustics, Speech and Signal Processing (ICASSP)},
  title={Muse: Multi-Modal Target Speaker Extraction with Visual Cues}, 
  year={2021},
  volume={},
  number={},
  pages={6678-6682},
  doi={10.1109/ICASSP39728.2021.9414023}}

@inproceedings{momuse,
author = {Li, Junjie and Zhang, Ke and Wang, Shuai and Lee, Kong Aik and Mak, Man-Wai and Li, Haizhou},
year = {2025},
month = {06},
pages = {1-6},
booktitle = {Proc. IEEE International Conference on Multimedia and Expo (ICME)},
title = {MoMuSE: Momentum Multi-modal Target Speaker Extraction for Real-time Scenarios with Impaired Visual Cues},
doi = {10.1109/ICME59968.2025.11209435}
}
\end{document}